\documentclass[%
 aip,
 amsmath,amssymb,
 reprint,%
]{revtex4-1}

\usepackage{graphicx}
\usepackage{dcolumn}
\usepackage{bm}

\usepackage[utf8]{inputenc}
\usepackage[T1]{fontenc}
\usepackage{mathptmx}
\usepackage{etoolbox}
\usepackage{hyperref}

\usepackage{subcaption}
\usepackage{mathtools}
\usepackage{afterpage}
\usepackage{upgreek}

\DeclareMathOperator*{\argmin}{argmin}

\graphicspath{{Figures/}}

\begin{document}

\setcounter{page}{0}
\afterpage{%
	\thispagestyle{empty}
    This article may be downloaded for personal use only. Any other use requires prior permission of the author and AIP Publishing.
    
    This article appeared in [Ngai \textit{et al.}, "Method of kinetic energy reconstruction from time-of-flight mass spectra", Rev. Sci. Instrum. 95, 033305 (2024)] and may be found at (\url{https://doi.org/10.1063/5.0201425}).
}
\clearpage
\clearpage

\makeatletter
\def\@email#1#2{%
 \endgroup
 \patchcmd{\titleblock@produce}
  {\frontmatter@RRAPformat}
  {\frontmatter@RRAPformat{\produce@RRAP{*#1\href{mailto:#2}{#2}}}\frontmatter@RRAPformat}
  {}{}
}%
\makeatother

\preprint{AIP/123-QED}

\title[Kinetic energy reconstruction from time-of-flight spectra]{Method of Kinetic Energy Reconstruction from Time-of-Flight Mass Spectra}
\author{A. Ngai}
\affiliation{Institute of Physics, University of Freiburg, Hermann-Herder-Str. 3, 79104 Freiburg, Germany}
\author{K. Dulitz}
\affiliation{Institut für Ionenphysik und Angewandte Physik, Universität Innsbruck, 6020 Innsbruck, Austria}
\author{S. Hartweg}
 \email{Sebastian.Hartweg@physik.uni-freiburg.de.}
\affiliation{Institute of Physics, University of Freiburg, Hermann-Herder-Str. 3, 79104 Freiburg, Germany}
\author{J. Franz}
\affiliation{Institut für Physik, Universität Kassel, Heinrich-Plett-Straße 40, 34132 Kassel, Germany}
\author{M. Mudrich}
\affiliation{Department of Physics and Astronomy, Aarhus University, Ny Munkegade 120, 8000 Aarhus C, Denmark}
\author{F. Stienkemeier}
\affiliation{Institute of Physics, University of Freiburg, Hermann-Herder-Str. 3, 79104 Freiburg, Germany}

\date{\today}

\begin{abstract}

We present a method for the reconstruction of ion kinetic energy distributions from ion time-of-flight mass spectra through ion trajectory simulations. In particular, this method is applicable to complicated spectrometer geometries with largely anisotropic ion collection efficiencies. A calibration procedure using a single ion mass peak allows the accurate determination of parameters related to the spectrometer calibration, experimental alignment and instrument response function, which improves the agreement between simulations and experiment. The calibrated simulation is used to generate a set of basis functions for the time-of-flight spectra, which are then used to transform from time-of-flight to kinetic-energy spectra. We demonstrate this reconstruction method on a recent pump-probe experiment by Asmussen \textit{et al.} (J. D. Asmussen \text{et al.}, \textit{Phys. Chem. Chem. Phys.}, \textbf{23}, 15138, (2021)) on helium nanodroplets and retrieve time-resolved kinetic-energy-release spectra for the ions from ion time-of-flight spectra.

\end{abstract}

\maketitle


\section{Introduction}
	Time-of-flight (TOF) spectroscopy is a simple yet powerful technique that is often used in parallel with other techniques in atomic and molecular physics, as well as in physical and analytical chemistry. Its application in mass spectrometry is widely used to disentangle products with different mass-to-charge ratios (m/q), which is crucial for example for the study of ionic dissociation or fragmentation processes. For example, in combination with photoionization or electron-impact ionization sources, TOF spectrometry can be used to monitor gas phase chemical reactions. Conditions that optimize the performance of TOF spectrometers such as temporal focusing and spatial focusing \cite{Wiley1955,Wollnik2013,Cai2015} have been extensively discussed, with the express purpose of increasing the resolution between different m/q. However, this optimization of mass resolution is detrimental for the retrieval of ion kinetic energy information, which is still implicitly encoded within the ion TOF spectra \cite{Riley_1994_IntJMassSpectrom}. There are ion spectrometers that are specialized in the determination of ion kinetic energies, such as velocity-map-imaging (VMI) spectrometers \cite{Eppink1997} and other purely TOF-based experimental techniques designed for the determination of ion kinetic energy distributions \cite{Riley_1994_IntJMassSpectrom,Ferreira2017,Ferreira2012,Cheng2018,Beavis_1991_ChemPhysLett}. It has even been possible to extract ion angular distributions based on TOF measurements alone \cite{Orgazolek_1998_JPhysChem, Hishikawa_1998_ChemPhys}.
	Widely-used Wiley-McLaren TOF mass spectrometers\cite{Wiley1955,Wollnik2013,Cai2015} allow the retrieval of ion momentum components parallel to the spectrometer axis, if operated under conditions not optimized for velocity focusing \cite{Tang2009,Garcia_2013_RevSciInstrum,Bodi2012,Hishikawa_1998_ChemPhys}.
	Assuming isotropic ion velocity distributions, the distribution of one momentum component is sufficient to retrieve a kinetic energy distribution. In some cases, Wiley-McLaren TOF spectrometers have been combined with imaging detectors to retrieve the remaining two momentum components of detected ions \cite{Tang2009,Garcia_2013_RevSciInstrum,Bodi2012}. 
	Reaction microscopes \cite{Ullrich2003,Dorner2000} also use position- and time-sensitive detectors to retrieve ion kinetic energy information in addition to the ion m/q and electron kinetic energies. In general, TOF spectrometers can be designed to provide a compromise between mass resolution, kinetic energy resolution and collection efficiency, which allows the possibility to create versatile instruments which can switch between different operational modes according to experimental needs.
	Experimental setups, using low-repetition-rate intense light sources, rather work with simple ion TOF spectrometers capable of handling higher maximum signal levels, often in combination with optimized photoelectron spectrometers \cite{Squibb2018,Braune_2016_JSynchrotronRadiat}.
	While the main purpose of these ion TOF spectrometers is the characterization of formed ions and ionic fragments according to their m/q, the respective ion kinetic energy releases can still be extracted by a careful analysis of the TOF peak shape and width \cite{Riley_1994_IntJMassSpectrom,Michiels_2021_JPhysB-AtMolOpt,Bapat_2006_IntJMassSpectrom,Fang_2012_PhysRevLett,Fang_2014_JPhysB-AtMolOptPhys,Boll_2016_StructDynam,Rupp_2016_PhysRevLett}. 
	The methods used to retrieve ion kinetic energy information from the data of such experiments and the assumptions behind them are, however, seldomly explained in much detail.


	In this article, we provide a general procedure to obtain kinetic-energy release (KER) distributions from ion TOF spectra through ion trajectory simulations, and discuss under which circumstances additional information can be obtained from a detailed structure of the kinetic energy distribution. An additional calibration procedure for the determination of the most relevant experimental parameters from a single mass peak will be introduced, and the uniqueness of this calibration procedure and the obtained kinetic energy distributions will be discussed. We first discuss the general outline for this KER reconstruction method, and then some strict requirements on which experimental parameters must be known beforehand for unique trajectory simulations.

	For illustration, we will explain in detail how this method was used for a previous experiment by Asmussen \textit{et al.} \cite{Asmussen2021} performed at the Low-Density-Matter (LDM) endstation of the FERMI free-electron laser in Trieste, Italy. This experiment examined the relaxation dynamics of excited states within helium nanodroplets. The helium nanodroplets were excited at the $1s3p/1s4p$ excitation band with an extreme ultraviolet (XUV) pump pulse with photon energy 23.7\,eV, followed by a delayed probe pulse of 3.2\,eV ionizing the droplets, in order to temporally resolve their relaxation dynamics. In addition to the previously extracted average ion kinetic energies, we now further complement the results by inferring the ion KER distributions for He$_{1,2,3}^{+}$. From this, we are able to resolve a multi-modal distribution for the He$^{+}$ ion signals with a high KE contribution, stemming from the ejection of excited Helium atoms (He$^{*}$) from the droplet, as opposed to a previous study which also observed a multi-model KE distribution induced by Interatomic Coloumbic Decay (ICD) leading to Coulomb explosion \cite{Shcherbinin2017,Sisourat_2021_NatPhys}. Although the involved ion spectrometer, operated in parallel to a magnetic bottle photoelectron spectrometer, was not designed to retrieve ion kinetic energies, we show that detailed ion kinetic energy distributions can be obtained and can contribute important information.




	\section{Time-of-flight from particle trajectories}

	TOF spectra are usually interpreted as mass-to-charge spectra. The conversion from TOF ($T$) to $m/q$ coordinates from a spectrometer with purely electrostatic fields is given by:
	\begin{equation}
	\label{QuadraticTOF_MassCharge}
		m/q=c T^2\ ,
	\end{equation}
	where $c$ is a proportionality constant. To determine $c$, some TOF peaks $T(m/q)$ corresponding to known mass-to-charge ratios are used. All ions in an arbitrary electrostatic potential $V({\bf{x}})$ follow the equation of motion $\ddot{{\bf{x}}}(t)=\frac{dV({\bf{x}})}{d{\bf{x}}} \Omega$, where $\Omega\coloneqq q/m$ is the charge-to-mass ratio. Let us assume we have a solution to that equation as a trajectory ${\bf{x}}(t)$. We can obtain another solution by appropriate re-scaling of ${\bf{x}}(t)$ and $\Omega$:
	\begin{equation}
	\label{ElectrodynamicSymmetry}
	\begin{aligned}
		t &\rightarrow kt\ , \\ 
		\Omega &\rightarrow k^2\Omega .
	\end{aligned}
	\end{equation}

	This is the basis of the robustness of TOF spectroscopy – if we only have electrostatic fields, then all particles, fulfilling some initial condition, follow the same spatial trajectory with some time-scaling. For a specific start time and a position in the trajectory, a time of flight $T_1$ taken by particle $1$ with charge-to-mass ratio $\Omega_1$, immediately determines the TOF $T_2$ of particle $2$ with charge-to-mass ratio $\Omega_2$, namely $T_2^2=\Omega_1/\Omega_2  T_1^2$, which leads to the quadratic relationship in Eq.~\eqref{QuadraticTOF_MassCharge}.

	For a one-dimensional trajectory of a single particle in a potential $V(x)$ with initial position $x$ and initial speed $v$, the flight time can be written as:
	\begin{equation}
	\label{TOF_Equation}
	\begin{aligned}
		T(x) = &\int_{x_b}^{x_f} \frac{dx'}{\sqrt{2\Omega [V(x_b)-V(x')]}} \\ &\pm \int_{x_b}^{x} \frac{dx'}{\sqrt{2\Omega[V(x_b)-V(x')]}}  ,
	\end{aligned}
	\end{equation}
	where $x_f$ is the spatial end point of the trajectory, $x_b$ is the turning-point where $V(x_b)-V(x)=v^2/2\Omega$ and $x_f\leq x\leq x_b$, and $v$ is the initial speed of the particle with charge-to-mass ratio $\Omega$. The $``\pm"$ denotes the starting conditions where $``-"$ has the particle’s initial velocity vector pointing towards the end-point, whereas the $``+"$ denotes backward-starting trajectories. Using this expression, ad-hoc analyses of quantities such as kinetic energy can already be performed \cite{Garcia_2013_RevSciInstrum,Hishikawa_1998_ChemPhys}. In the case of three-dimensional trajectories, there is no analogous simple expression due to integrability, although the above equation may still work for some spectrometer geometries to retrieve the momentum along the spectrometer axis \cite{Garcia_2013_RevSciInstrum}. 

	In the end, we want to consider the flight times of a collection of particles flying in an arbitrary 3-dimensional spectrometer geometry, which may include magnetic fields. For a known spectrometer geometry, we can simulate the spatially- and velocity-dependent TOF map $T({\bf{x}},{\bf{v}};\Omega)$. If we know the initial spatial/velocity distribution of the ions $G({\bf{x}}, {\bf{v}}; \Omega)$ characterized by initial position ${\bf{x}}$, initial velocity ${\bf{v}}$, and charge-to-mass ratio $\Omega$ we can calculate the corresponding time-of-flight spectrum $f(t)$ where $t$ is the TOF coordinate as:
	\begin{equation}
	\label{TOF_Transformation}
		f(t)=\sum_{\Omega} \int_\Phi d^3{\bf{x}}\ d^3{\bf{v}}\ G({\bf{x}},{\bf{v}};\Omega) \delta\left(t-T\left({\bf{x}},{\bf{v}};\Omega\right)\right) ,
	\end{equation}
	where $\delta(x)$ is the Dirac delta distribution, and $\Phi$ is the phase-space volume of the ions. This is equivalent to a binning procedure for a discrete sampling scheme. Note that Eq.~\eqref{TOF_Transformation} is analogous to a density of states formalism. This integral can be numerically sampled through a simple binning procedure for discrete values of $t$. Otherwise, Eq.~\eqref{TOF_Transformation} can be further simplified \cite{Hoermander2003} as:
	\begin{equation}
	\label{TOF_TransformationAnalytic}
		f(t) = \sum_{\Omega} \int_{\partial \Phi_t} dS_t \ \frac{G({\bf{x}},{\bf{v}}; \Omega)}{|{\bf{\nabla}}T({\bf{x}},{\bf{v}}; \Omega)|} ,
	\end{equation}
	where $\partial \Phi_t$ is the 2D (possibly disjointed) surface with $t=T({\bf{x}},{\bf{v}};\Omega)$.

	The task now is to invert Eq.~\eqref{TOF_Transformation} to determine $G({\bf{x}},{\bf{v}};\Omega)$ given $f(t)$ and $T({\bf{x}},{\bf{v}};\Omega)$. In the following section, we will show how this information can be used to obtain mass, charge, and initial velocities from TOF data.

	\section{Reconstruction through basis functions}

	In its most basic form, we consider the forward transformation from the KER spectrum ${\bf{g}}$ to the TOF spectrum ${\bf{f}}$ as an analogous linear matrix equation
	\begin{equation}
	\label{basic_linear_system}
		{\bf{f}}={\bf{K}} {\bf{g}} ,
	\end{equation}
	where we make the following identifications between Eq.~\eqref{TOF_Transformation} and  Eq.~\eqref{basic_linear_system}:
    \begin{subequations}
    \begin{eqnarray}
        f(t) &\rightarrow {\bf{f}}, \label{eq:discrete_TOF_spectrum}
        \\
		\sum_{\Omega} \int_\Phi d{\bf{x}}d{\bf{v}}\ \delta(t-T({\bf{x}},{\bf{v}};\Omega)) &\rightarrow {\bf{K}}, \label{eq:discrete_TOFMapping}
        \\
		G({\bf{x}},{\bf{v}};\Omega) &\rightarrow {\bf{g}}. \label{eq:discrete_source}
    \end{eqnarray}
	\end{subequations}
	By appropriately choosing the sizes of the TOF and KER basis sets, i.e.\ the discretization of Eqs.~\eqref{eq:discrete_TOFMapping} and ~\eqref{eq:discrete_source}, the  matrix ${\bf{K}}$ is designed to be square and we will assume its invertibility. We could obtain ${\bf{g}}$ as the direct inversion solution ${\bf{g}}_D$:
	\begin{equation}
		{\bf{g}}_{D}={\bf{K}}^{-1} {\bf{f}} .
	\end{equation}

	However, the presence of noise may yield nonphysical results, especially  if ${\bf{K}}$ is near-singular. To circumvent this, we instead use a least-squares method to find the minimal least-squares solution ${\bf{g}}$, and relax the assumption of the invertibility of ${\bf{K}}$ to the new assumption that ${\bf{K}}{\bf{K}}^T$ is invertible.
	\begin{equation}
	    \label{LeastSquaresMinimization}
		{\bf{g}}_{LS} \coloneqq \argmin_{{\bf{g}}} \left\{\left\|{\bf{f}}-{\bf{K}}{\bf{g}}\right\|_2\right\} .
	\end{equation}
	This least-square solution ${\bf{g}}_{LS}$ is obtained through the matrix pseudoinverse ${\bf{M}}$ (Moore-Penrose pseudoinverse):
	\begin{equation}
	    \label{MoorsePenrosePseudoinverse}
		{\bf{g}}_{LS}={\bf{M}}{\bf{f}}={\bf{K}}^T ({\bf{K}}{\bf{K}}^T)^{-1} {\bf{f}} .
	\end{equation}

	There is yet another potential issue; with a large-enough basis set, we may encounter the problem of over-fitting if the matrix ${\bf{K}}$ is near-singular. This over-fitting can be compensated by replacing the original inversion problem with an approximate problem which is well-posed. We do this through a regularization scheme -- on top of the least-squares constraint, we minimize the $L_2$-norm of ${\bf{g}}$ where the weight between the least-squares or norm constraint is determined by an arbitrary choice of regularization parameters $\Lambda_0, \Lambda_1$:
	\begin{equation}
	    \label{eq:TikhanovCondition}
	    {\bf{g}}_{R} \coloneqq \argmin_{{\bf{g}}} \left\{\left\|{\bf{f}}-{\bf{K}}{\bf{g}}\right\|_2 + \Lambda_0\left\|{\bf{g}}\right\|_2 + \Lambda_1\left\|{\bf{D}}{\bf{g}}\right\|_2\right\} ,
	\end{equation}
	where $\bf{D}$ is the first-order finite difference matrix, and $\Lambda_0$ and $\Lambda_1$ are often known as Tikhanov parameters. This regularized solution ${\bf{g}}_{R}$ is obtained through the regularized inverse ${\bf{T}}$ where:
	\begin{equation}
	    \label{eq:TikhanovMatrix}
	    {\bf{g}}_{R}={\bf{T}}{\bf{f}}={\bf{K}}^T \left({\bf{K}}{\bf{K}}^T -\Lambda_0 {\bf{I}}-\Lambda_1 {\bf{D}}^2 \right)^{-1} {\bf{f}} .
	\end{equation}
	As regularization is a process which necessarily introduces bias, care must be taken in choosing appropriate regularization parameters such that this systematic bias does not dominate over the underlying physical results (see Appendix A in the Supplementary Material).

	\section{Instrument function}

\begin{figure}[t]
        \centering
        \begin{subfigure}[t]{\linewidth}
    		\centering
    		\includegraphics[width=\linewidth*4/4]{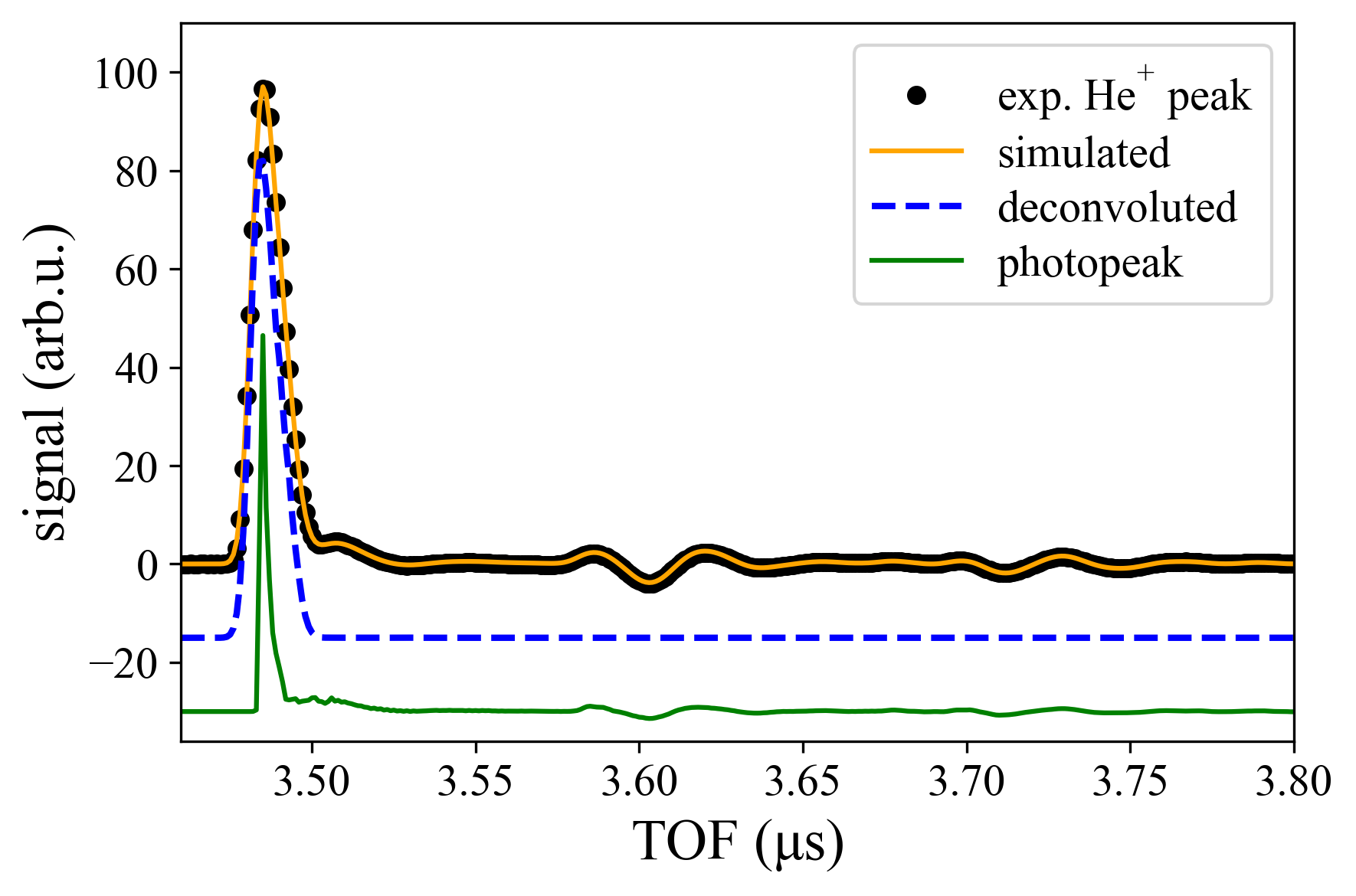}
            \caption[subfig:CalibrationPeak]{\label{subfig:CalibrationPeak}}
        \end{subfigure}
        
        \begin{subfigure}[t]{\linewidth}
    		\centering
    		\includegraphics[width=\linewidth*4/4]{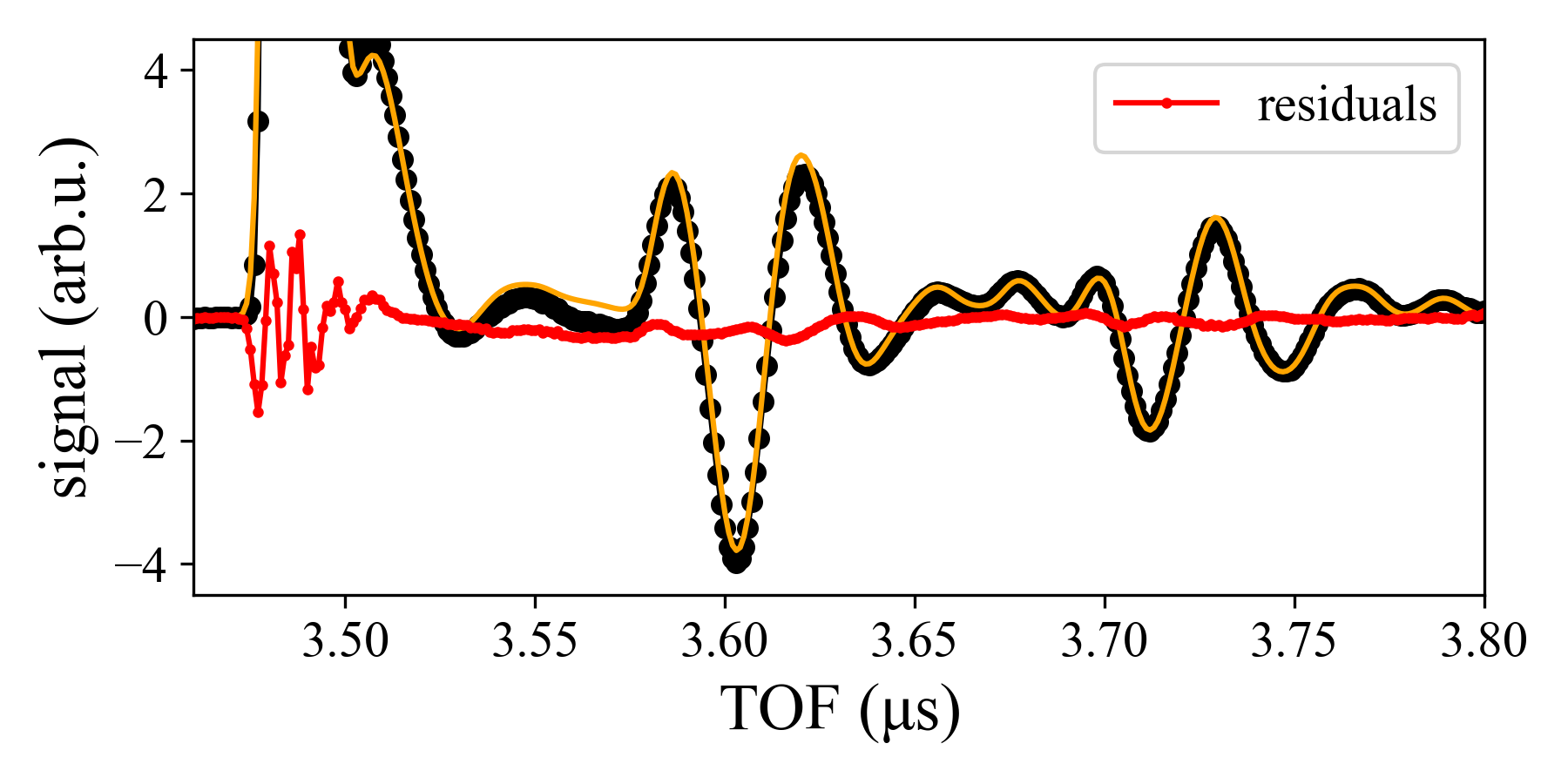}
            \caption[subfig:CalibrationResiduals]{\label{subfig:CalibrationResiduals}}
        \end{subfigure}
		\caption[fig:ResponseFunctionConvolution]{\label{fig:ResponseFunctionConvolution}(\subref{subfig:CalibrationPeak}) The experimental He$^+$ TOF peak for a zero kinetic energy atomic beam (dots) and its simulated counterpart (orange line) based on the scattered light signal (green line, temporally shifted for comparison). The deconvoluted spectrum (dashed blue line) is also shown. The blue and green lines are offset for visibility. (\subref{subfig:CalibrationResiduals}) Zoomed-in view with residuals (red line).}
	\end{figure}

	Up to now, we have theoretically described how the TOF spectrum $f(t)$ is derived from the properties of the spectrometer and initial spatial and velocity distribution of the particles. In a realistic application, the detector response modifies $f(t)$, so that what we observe as the spectrum is not $f(t)$ but rather $f_{\text{obs}}(t)$:
	\begin{equation}
	\label{eq:InstrumentConvolution}
		 f_{\text{obs}}(t) = f(t) * h(t) ,
	\end{equation}
	where the device-specific instrument function $h(t)$ is convoluted with the real signal $f(t)$ (see Fig.\ \ref{fig:ResponseFunctionConvolution}). $h(t)$ may be observed as a ringing response (e.g.\ of a multichannel-plate (MCP) detector).

	There is no universal rule governing the form of $h(t)$, and ad-hoc methods may be necessary. A form for $h(t)$ we typically find is:
	\begin{equation}
	\label{eq:ResponseFunction}
		h(t) = h_{\text{photo}}(t) + c_\delta \delta(t) + c_\tau \left(e^{-t/\tau_g} * h_{\text{photo}}(t) \right) ,
	\end{equation}
	where ``$*$'' denotes convolution, and $h_{\text{photo}}(t)$ is the signal created by scattered light on the MCP detectors which typically has characteristic ringing features (see Fig.\ \ref{fig:ResponseFunctionConvolution}). Ideally, $h_{\text{photo}}(t)$ is the exact impulse response of the detector, but the scattered light peak is the closest approximation in our case. The second term $c_\delta(t)$ compensates for under-sampling at the sharp scattered light signal, by adjusting the relative intensities of the sharp signal to the ringing response. The third term in the expression above, with $c_\tau<0$, describes an overshoot of the signal level to negative values, eventually decaying back to zero with a time constant $\tau_g$.

	In comparison, for an ideal detector with no ringing response, $h(t)=\delta(t)$ would be the corresponding response function, i.e.\ the observed TOF spectrum $f_{\text{obs}}(t) =f(t)$ is the same as the actual ion TOF spectrum.

	\section{Application to experimental ion TOF spectra}

	The forms of the initial spatial distribution $G({\bf{x}},{\bf{v}};\Omega)$ and TOF map $T({\bf{x}},{\bf{v}};\Omega)$ are specific to different experimental geometries, and we now consider the crossed-beam geometry in Ref.\ \cite{Asmussen2021}. In our case, we first make the following simplifications of an ion distribution whose velocity-component is spatially-independent, and is isotropic in velocity:
	\begin{equation}
	    G({\bf{x}},{\bf{v}};\Omega) = \mu({\bf{x}})\rho(|{\bf{v}}|;\Omega) .
	\end{equation}
	The first simplification $G({\bf{x}},{\bf{v}};\Omega) = \mu({\bf{x}})\rho({\bf{v}};\Omega)$ is trivial if initial ion velocities are insignificant. This is true for ionization events that do not lead to fragmentation, due to momentum/energy conservation. When significant velocities do exist, then this simplification requires that the kinetic energy release of ions is mostly insensitive to the spatial profile of the laser or another ionization source. In the case of a process where ions are produced by single-photon ionization, this requirement is clearly fulfilled. However, when there are several processes involving a different number of photons producing ions of the same mass but different kinetic energies, this requirement may not be fulfilled.

	In our experiments, we excited the helium nanodroplets with a single XUV photon and subsequently probed the dynamics using one- and two-photon ultraviolet (UV) ionization. Due to the low number of photons involved, this single-photon ionization assumption is likely a good approximation and we do not expect a spatial dependence of the kinetic energy within the ionization volume. In the case of Coulomb explosion from molecular fragmentation in the gas phase or a nanoplasma, this simplification is justified.

	The second simplification $\rho({\bf{v}};\Omega)=\rho(v;\Omega)$ with $v=|{\bf{v}}|$ assumes that the initial ion velocities are directionally isotropic. In the case of ions originating from within a cluster where collisions are expected, this is a good assumption. For ionic dissociation reactions of small molecules caused by interactions with polarized light, this assumption may not be generally fulfilled. 

	In this form, the question of converting TOF to KER spectra is framed as determining the isotropic velocity distribution $\rho(|{\bf{v}}|;\Omega)$. To do this, we divide our approach into three steps:
	\begin{enumerate}
	    \setlength\itemsep{0em}
	    \item determining $\mu({\bf{x}})$ by calibrating and verifying simulation parameters, using a TOF signal created by ions of zero kinetic energy,
	    \item expanding $\mu({\bf{x}})\rho_i(v;\Omega)$ as a set of basis functions labelled by $i$, and
	    \item determining $\rho(v;\Omega)$ and hence KER spectra by fitting the basis functions to TOF spectra.
	\end{enumerate}
	The geometry of the experiment is shown in Fig.\ \ref{fig:SpectrometerGeometry} (detailed cut through part of the spectrometer is depicted in Appendix B in the Supplementary Material). Importantly, this spectrometer necessarily differs from traditional Wiley-McLaren designs, by including a permanent magnet. The ion extraction field is pulsed shortly after the ionizating laser pulse, deviating from optimal time-focusing conditions, in order to not affect the electron detection placed opposite to the ion spectrometer \cite{Squibb2018}.

	\begin{figure}[t]
	    \centering
	    \includegraphics[width=\linewidth*3/4]{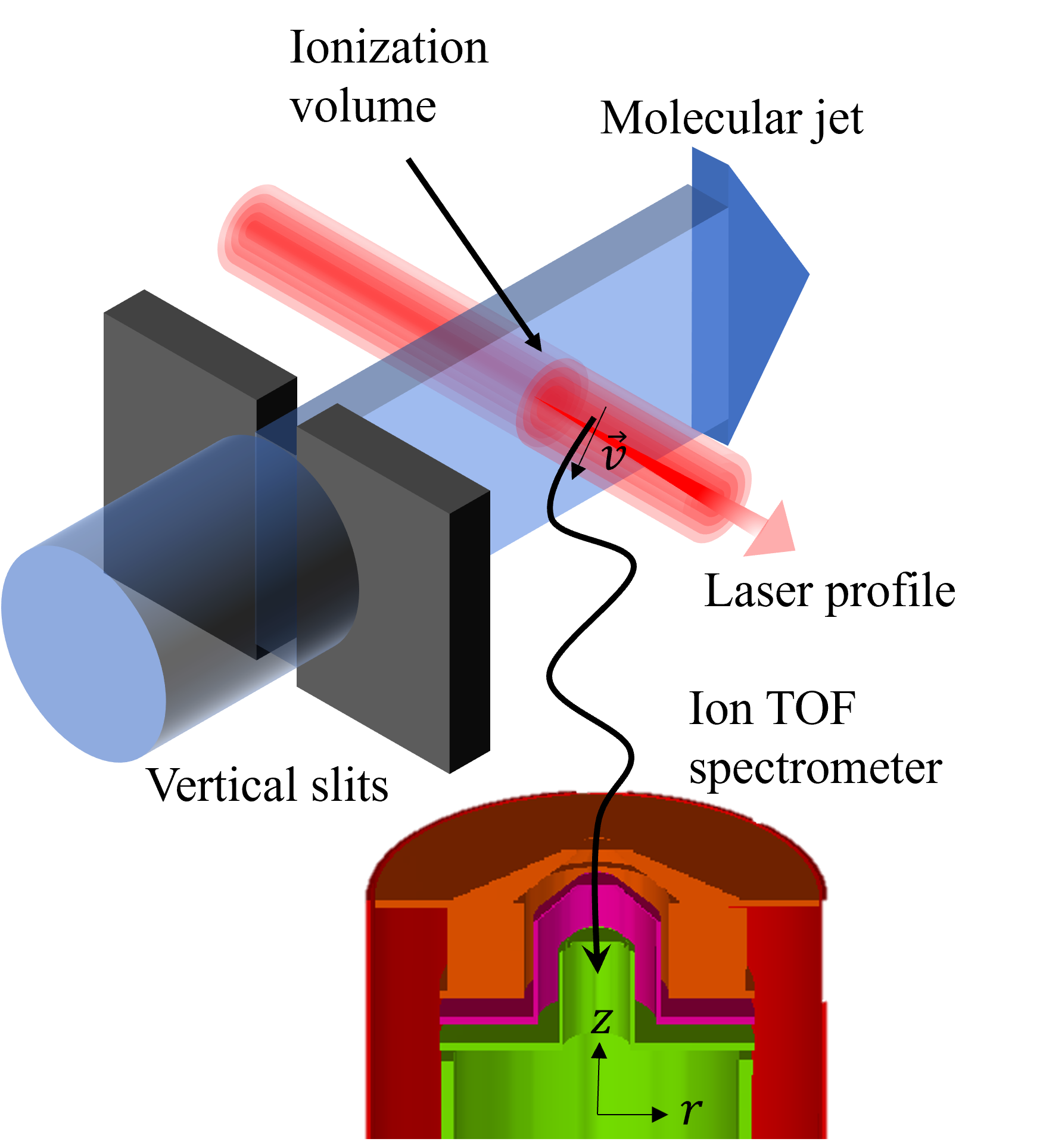}
	    \caption[SpectrometerGeometry]{\label{fig:SpectrometerGeometry} Ionization volume $\mu({\bf{x}})$ formed by the overlap of the laser profile and gas jet. Ions are extracted from the ionization volume into the cylindrically-symmetric ion TOF spectrometer and impinge on an MCP. Colours for the ion spectrometer correspond to different electrode potentials.}
	\end{figure}

\begin{figure}[t]
		\centering
		\setlength{\tabcolsep}{0pt}
		\begin{tabular}[t]{c c}
			\begin{subfigure}{0.5\linewidth}
				\centering
				\includegraphics[width=\linewidth*4/4]{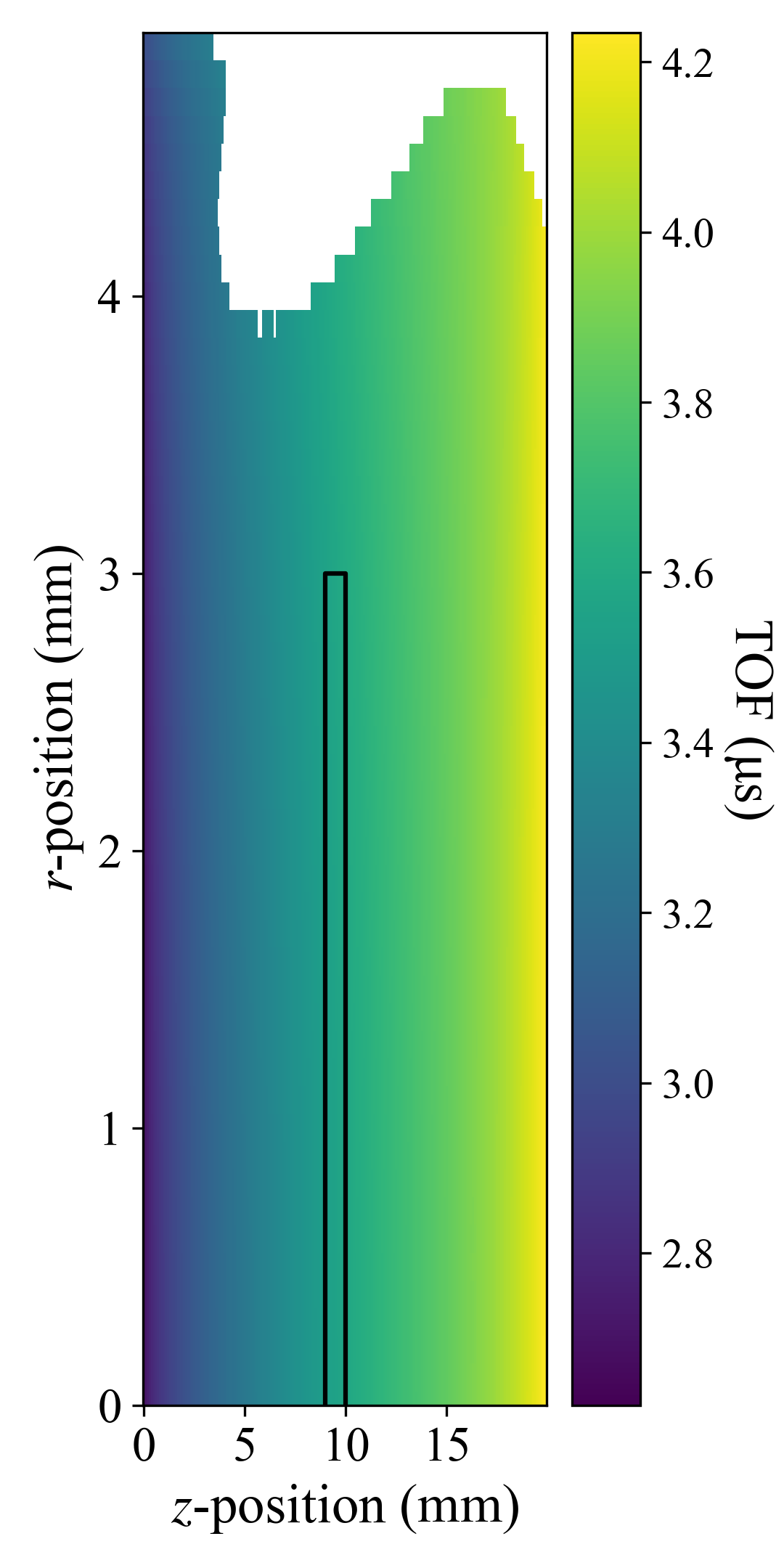}
				\caption[TOF_map_2D]{\label{TOF_map_2D} }
			\end{subfigure}
			&
			\begin{tabular}{c}
				\begin{subfigure}[t]{0.5\linewidth}
					\centering
					\includegraphics[width=\linewidth*4/4]{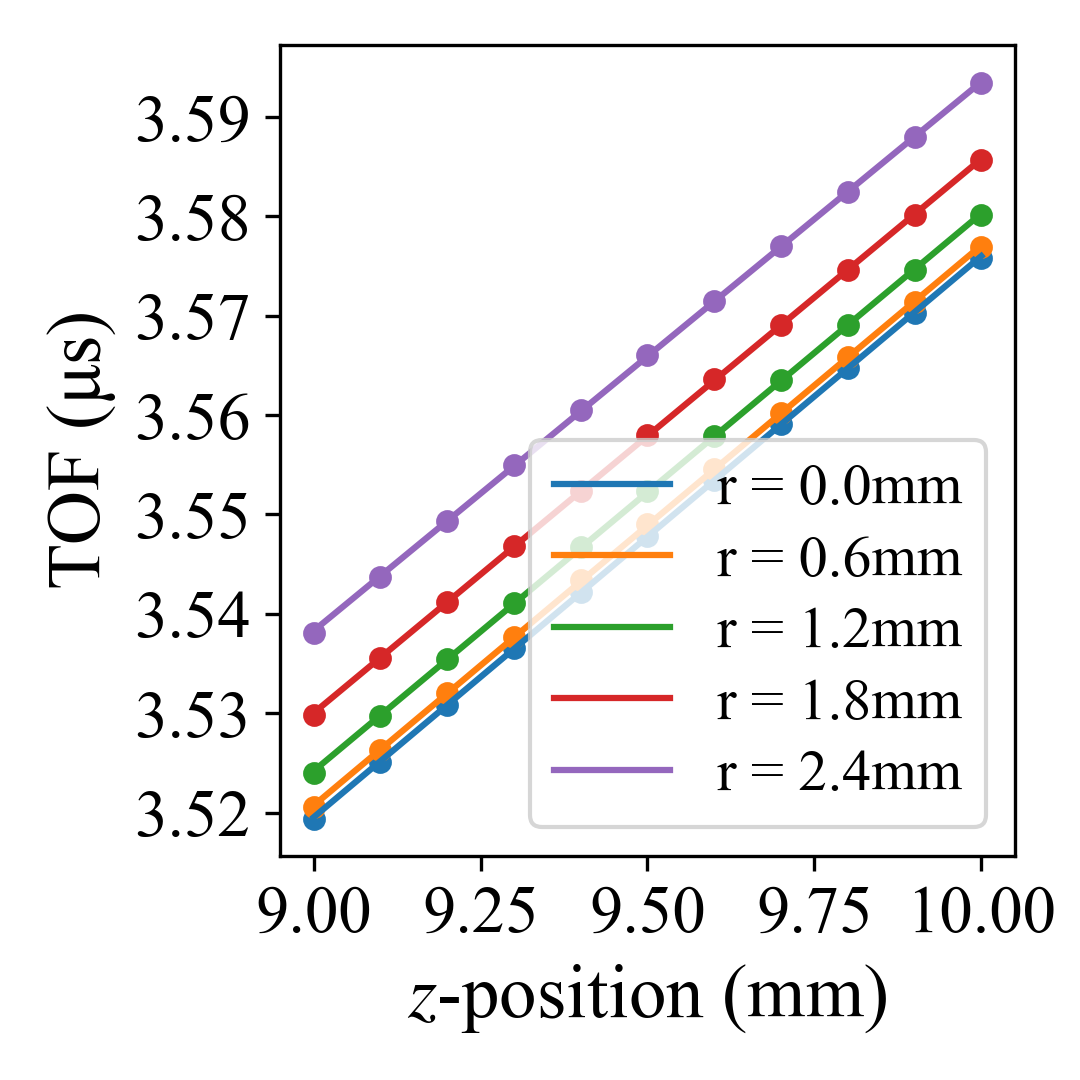}
				\caption[TOF_map_linear_z]{\label{TOF_map_linear_z} }
				\end{subfigure}\\
				\begin{subfigure}[t]{0.5\linewidth}
					\centering
					\includegraphics[width=\linewidth*4/4]{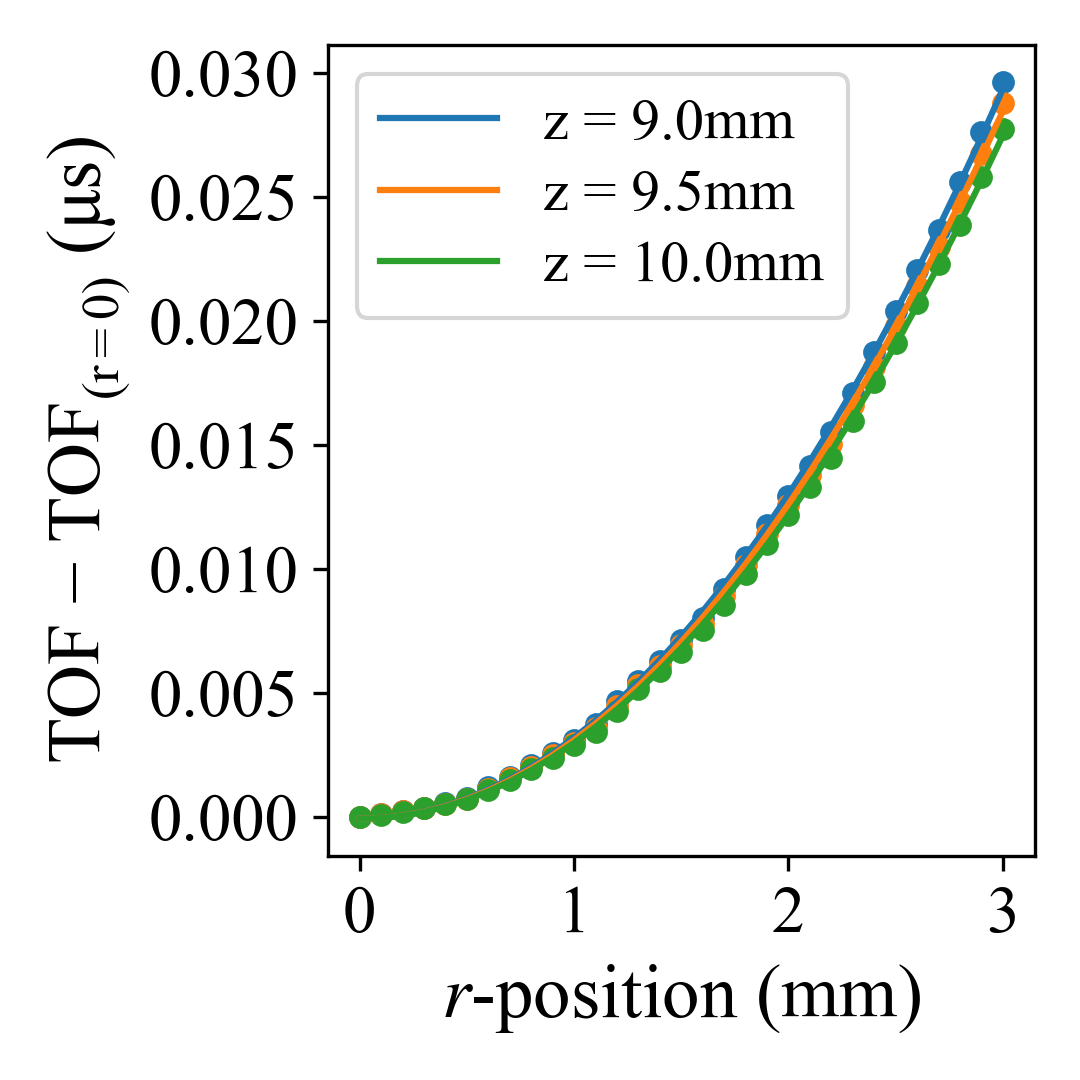}
				\caption[TOF_map_quadratic_r]{\label{TOF_map_quadratic_r} }
				\end{subfigure}
			\end{tabular}\\
		\end{tabular}
		\caption[TOFMap]{\label{TOFMap} TOF map as interpreted in Eq.~\eqref{eq:TOF_TaylorCylindrical}. (\subref{TOF_map_2D}) TOF from the entire extraction region. The area enclosed by the black rectangle denotes the ionization volume between the laser and gas jet. (\subref{TOF_map_linear_z}) Linear dependence  of the TOF on the initial axial position. (\subref{TOF_map_quadratic_r}) Quadratic dependence of the TOF on the initial radial position.}
	\end{figure}

	First, we assume the form for the ionization volume $\mu({\bf{x}})$ based on our specific experimental geometry in Fig.\ \ref{fig:SpectrometerGeometry} with the set of spatial parameters $P_{\mu} \coloneqq \{x_L,z_L,\sigma_L,y_G,\sigma_G \}$:
	\begin{subequations}
    \label{SpatialLaserGas}
    \begin{gather}
		\mu({\bf{x}}) = I_L({\bf{x}}) I_G({\bf{x}}) ,\label{eq:SpatialOverlap}
        \\
		I_L({\bf{x}}) = \frac{1}{\sigma_L^2 2\pi} e^{-\frac{1}{2}(\frac{z-z_L}{\sigma_L})^2} e^{-\frac{1}{2}(\frac{x-x_L}{\sigma_L})^2}, \label{eq:LaserProfile}
        \\
		I_G({\bf{x}}) = \frac{1}{2 \sigma_G} H\left(\sigma_G - |y-y_G|\right), \label{eq:GasProfile}
    \end{gather}
	\end{subequations}
	where $I_L(r,z,\theta)$ is the Gaussian laser profile, and $I_G(r,z,\theta)$ is the rectangular gas jet profile.
	Second, we simulate the TOF map $T(r,z)$ for ions with zero-initial velocity (Fig.\ \ref{TOFMap}) by running particle trajectories with the program SIMION \cite{SIMION}. Note that it is well approximated by a Taylor expansion up to the linear term in z and quadratic term in $r$:
	\begin{equation}
		\label{eq:TOF_TaylorCylindrical}
	    \begin{aligned}
		T(r,z)&\approx A_{00}+A_{01}z+A_{10}r^2+A_{11}zr^2.
	    \end{aligned}
	\end{equation}
	Now we combine the ionization volume and the TOF map to get TOF spectral functions. In this case, Eq.~\eqref{TOF_TransformationAnalytic} has the explicit form:
	\begin{equation}
	\label{TOF_TransformationCylindrical}
		\begin{aligned}
		f(t) &=2\pi \int_0^\infty \!\! dr \left\{r\ \mu(r,z)\left|\frac{\partial T(r,z)}{\partial z }\right|^{-1}\right\}_{(z(r,t):t=T)}.
		\end{aligned}
	\end{equation}
	To get the observed ion signal $f_{\text{ion}}$, we convolve $f(t)$ with the instrument function described previously with Eqs.~\eqref{eq:InstrumentConvolution} and \eqref{eq:ResponseFunction}, which contain the explicit instrument parameters $P_{h} \coloneqq \{h_{\text{photo}}, c_{\delta},c_{\tau},\tau_g\}$.

	To determine $\mu(r,z)$ up to the spatial parameters $P_{\mu}$ in Eq.~\eqref{SpatialLaserGas}, i.e.\ spatial positions and widths of the crossed laser beam and gas jet, and the instrument parameters $P_{h}$ in Eq.~\eqref{eq:ResponseFunction} (see Fig.\ \ref{fig:ResponseFunctionConvolution}), we did a simultaneous fitting routine using He ions with vanishing kinetic energy . Ignoring this ringing effect i.e.\ the instrument function, leads to an overestimation in the ionization volume (50 $\upmu$m) compared to the nominal value (30 $\upmu$m), and compared to including the ringing effect (45 $\upmu$m), as the calibration procedure is very sensitive to the shape of a zero-velocity TOF peak.

	We compared the convolved simulated TOF spectrum $f(t)$ with our reference TOF spectrum (Fig.\ \ref{fig:ResponseFunctionConvolution}) and minimized the least-squares difference within this routine. The resulting fit parameters which define the ionization volume are given in Table \ref{table:FitRoutineParameters}.

	\setlength{\tabcolsep}{6pt}
	\renewcommand{\arraystretch}{1} 
	\begin{table}[b!]
		\centering
		\caption[table:FitRoutineParameters]{\label{table:FitRoutineParameters} Fit parameters of the ionization volume according to Eqs.~\eqref{eq:ResponseFunction} and \eqref{eq:SpatialOverlap}, found through the calibration procedure.}
    \begin{ruledtabular}
		\begin{tabular}{ l | c | c } 
			Description (\& units)  & Fit value & Error (1$\sigma$) \\
			Variable in Eqs.~\eqref{eq:ResponseFunction},\eqref{eq:SpatialOverlap} &  &   \\ 
			
			\hline\hline
			x-centre of laser ($\upmu$m)   & 8374.65  & 0.11( + 350)\footnotemark[1] \\
			\quad 0 $\leq x_L \leq$ 20000  &          &                    \\		
			z-centre of laser (mm)         & 0        & fixed\footnotemark[2]     \\
    		\quad 0 $\leq z_L < ~$5        &          &                    \\	
			width of laser ($\upmu$m)      & 44.34    & 0.11               \\	
    		\quad 0 $< \sigma_L$           &          &                    \\			
			\hline
			y-centre of gas jet (mm)       & 0.12     & 0.06               \\
    		\quad 0 $\leq y_G < ~$5        &          &                    \\		
			width of gas jet (mm)          & 1.86     & 0.06               \\
    		\quad $\sigma_G < ~$2.5        &          &                    \\			
			\hline	
			photopeak undersampling        & 1.865    & 0.028              \\
    		\quad 0 $< c_\delta$           &          &                    \\					
			baseline decay (ns)            &  9.67    & 0.24               \\
    		\quad 0 $< \tau_g$             &          &                    \\				
			baseline factor                &  -0.0216 & 0.0009             \\
    		\quad $c_\tau \leq$ 0          &          &                    \\				
		
		\end{tabular}
    \end{ruledtabular}
    \footnotetext[1]{Additional error from the uncertainty of the electrode voltages (see Appendix C in the Supplementary Material).}
    \footnotetext[2]{The assumption $z_L=0$ was done for practical reasons, but otherwise did not significantly affect the fitting.}
		
	\end{table}

	This initial calibration procedure automatically takes into account any spatial broadening e.g.\ from a time-delayed extraction, as well as the effect of the permanent magnetic field. Any initial thermal velocities e.g.\ from the gas jet are on the order of a few meV and are neglected. Further, note that only a calibration to a single atomic peak (e.g.\ He$^{+}$) is needed; corollarily this uniquely fixes the trajectories of every other ion (e.g.\ He$_{2,3}^{+}$).

\begin{figure}[t]
	    \centering
	    \includegraphics[width=\linewidth*9/10]{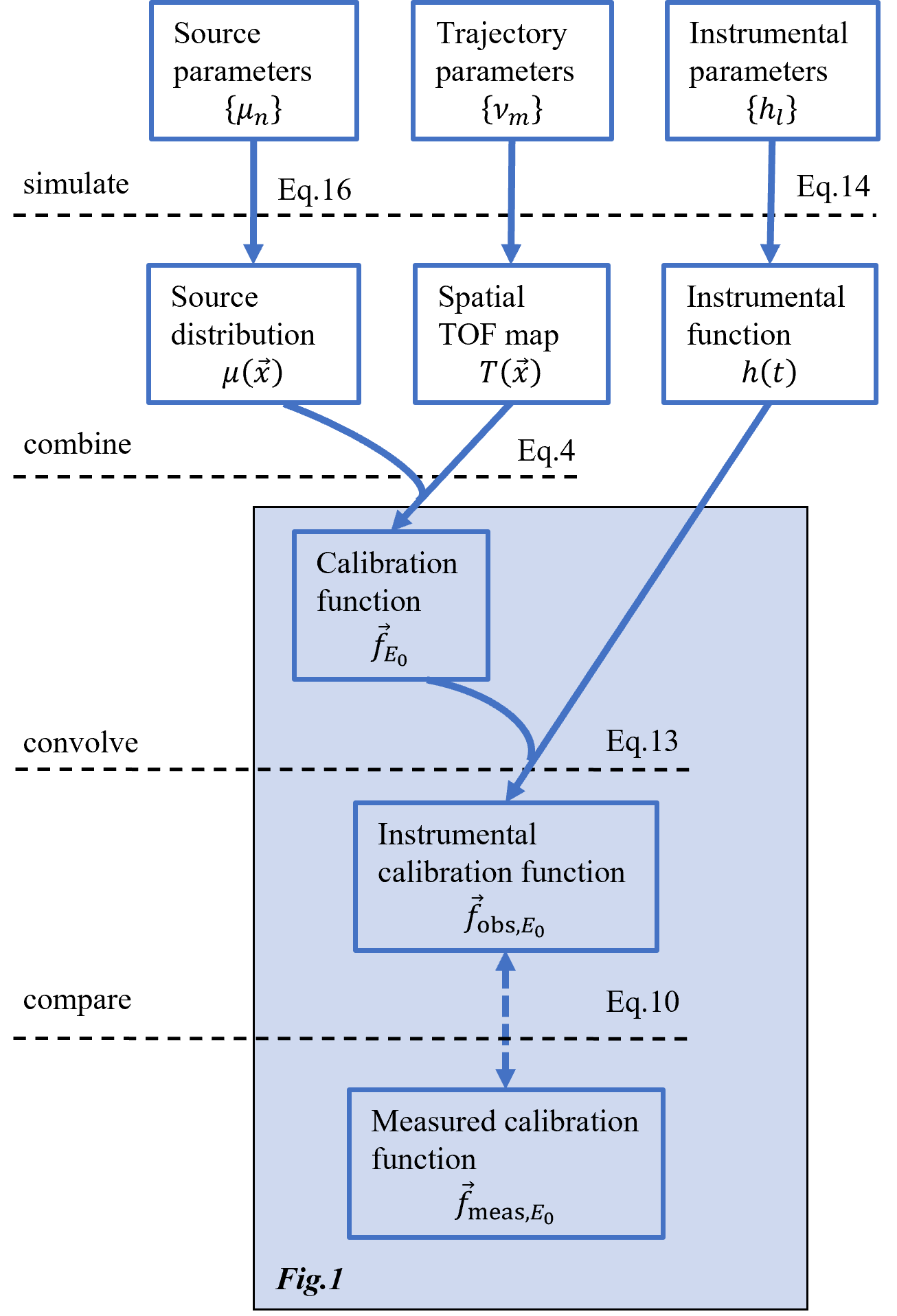}
	    \caption{Flowchart for the calibration stage. Four steps are involved: simulation of source parameters (Eq.~\eqref{eq:SpatialOverlap}), zero-initial-KE trajectories, and instrument response (Eq.~\eqref{eq:ResponseFunction}); combining the source and trajectory simulations (Eq.~\eqref{TOF_Transformation}); convolving the simulated spectrum with the instrument function (Eq.~\eqref{fig:ResponseFunctionConvolution}); and a least-squares comparison between the resulting spectrum and an experimental spectrum.}
	    \label{fig:CalibrationStepsFlowchart}
	\end{figure}

	We give a graphical summary of this calibration procedure in Fig.\ \ref{fig:CalibrationStepsFlowchart}.

\begin{figure}
		\centering
		\includegraphics[width=\linewidth*4/4]{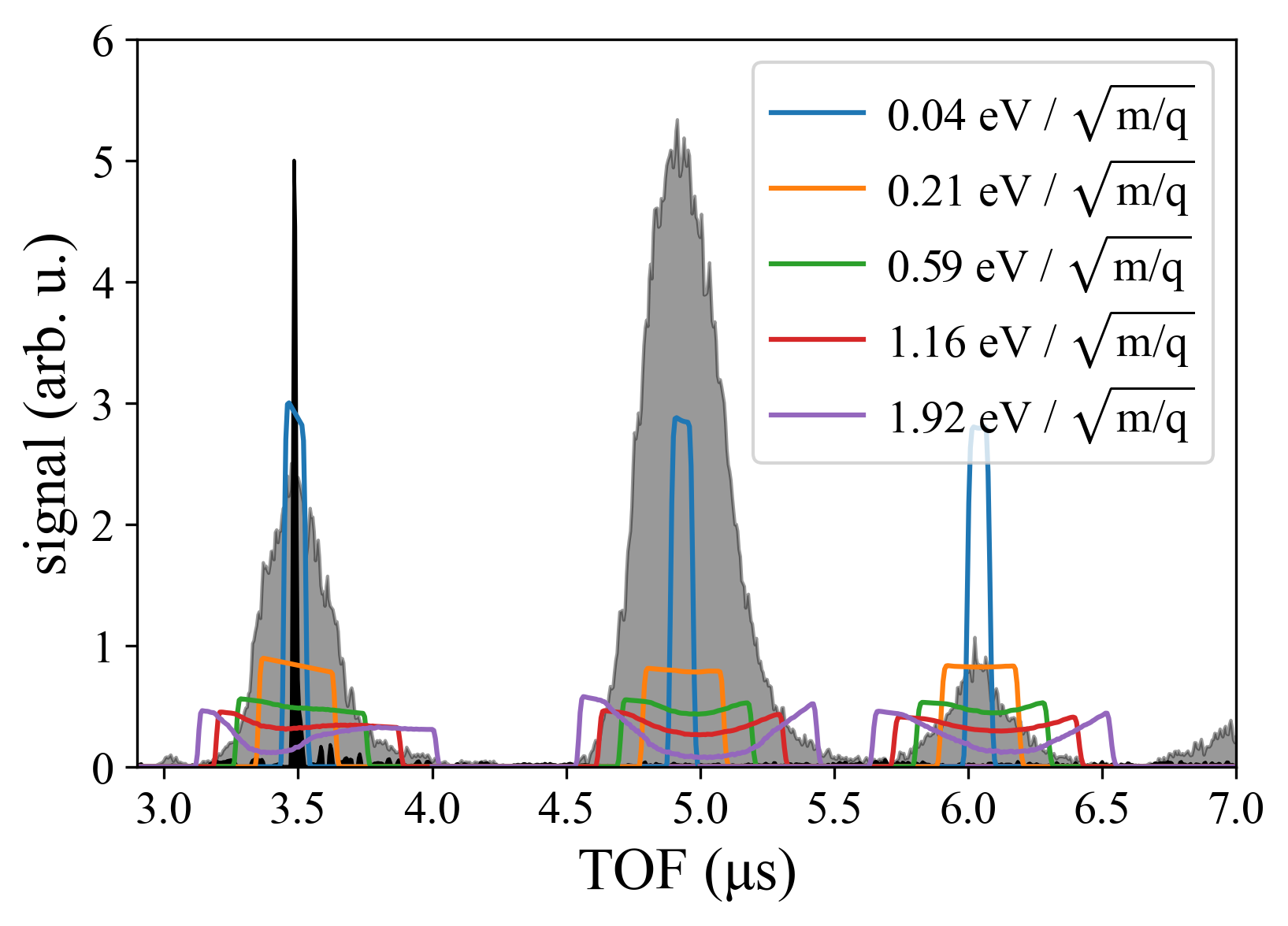}
		\caption[fig:TOFBasisFunctions]{\label{fig:TOFBasisFunctions} Typical atomic TOF spectra (black-shaded region) are very narrow and only contain He$^{+}$ centered around 3.5\ $\mu$s. Typical cluster TOF spectra (grey-shaded region) are much broader and contain He$_{2}^{+}$ and He$_{3}^{+}$ centered around 4.9, 6.0\ $\mu$s respectively, in addition to He$^{+}$. TOF basis functions with similar widths and initial velocities between He$^{+}_{1,2,3}$ are shown with the same colour. Their corresponding energy is dependent on their m/q, and can be found in the legend.}
	\end{figure}

	With the simulation calibrated with zero initial velocity ions, we then simulate additional trajectories with different initial velocities i.e., we repeat the simulations in Fig.\ \ref{TOF_map_2D} for each velocity vector. From these new trajectories, we build energy-dependent basis functions through a binning procedure (Eq.~\eqref{TOF_Transformation}); a few example basis functions are shown in Fig. \ref{fig:TOFBasisFunctions}. 

	We used a Tikhonov-regularized inverse for the reconstruction (Eq.~\eqref{eq:TikhanovMatrix}), with the regularization parameter as small as possible while still retaining peaks that are above the noise threshold (results for various regularization parameters are shown in Appendix A in the Supplementary Material).

\begin{figure*}[t]
		\centering
		\begin{subfigure}[t]{0.33\linewidth}
			\centering
			\includegraphics[width=\linewidth*4/4]{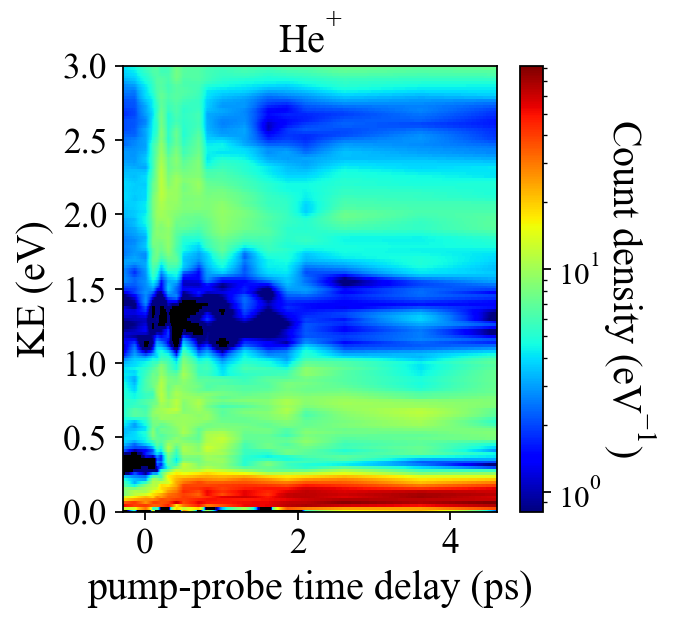}
			\caption[subfig:PlaneGasKER_He+]{\label{subfig:PlaneGasKER_He+} }
		\end{subfigure}%
		~
		\begin{subfigure}[t]{0.33\linewidth}
			\centering
			\includegraphics[width=\linewidth*4/4]{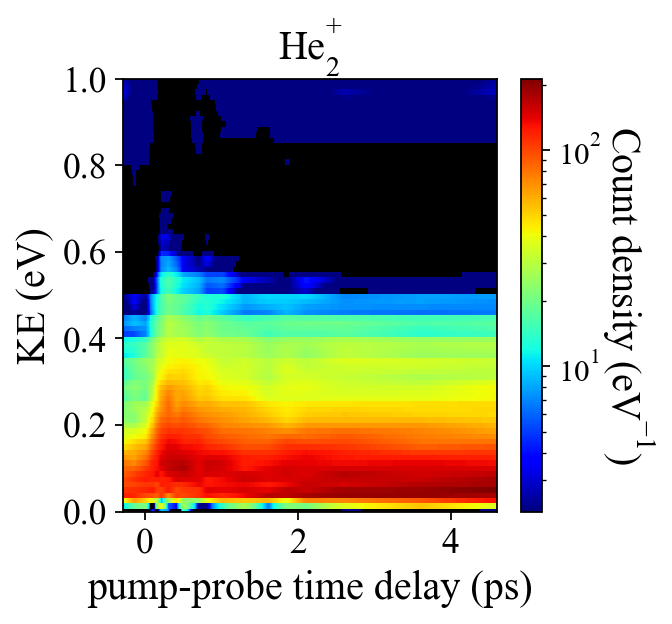}
			\caption[subfig:PlaneGasKER_He2+]{\label{subfig:PlaneGasKER_He2+} }
		\end{subfigure}%
		~
		\begin{subfigure}[t]{0.33\linewidth}
			\centering
			\includegraphics[width=\linewidth*4/4]{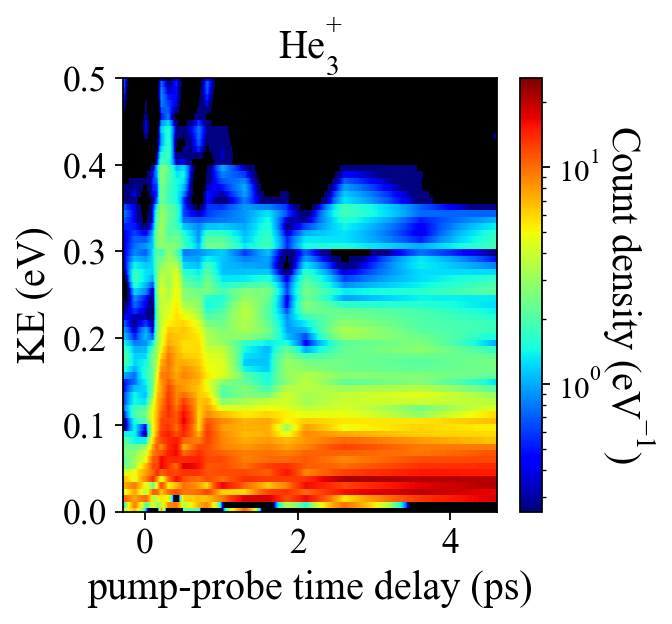}
			\caption[subfig:PlaneGasKER_He3+]{\label{subfig:PlaneGasKER_He3+} }
		\end{subfigure}%

		\begin{subfigure}[t]{0.47\linewidth}
			\centering
			\includegraphics[width=\linewidth*4/4]{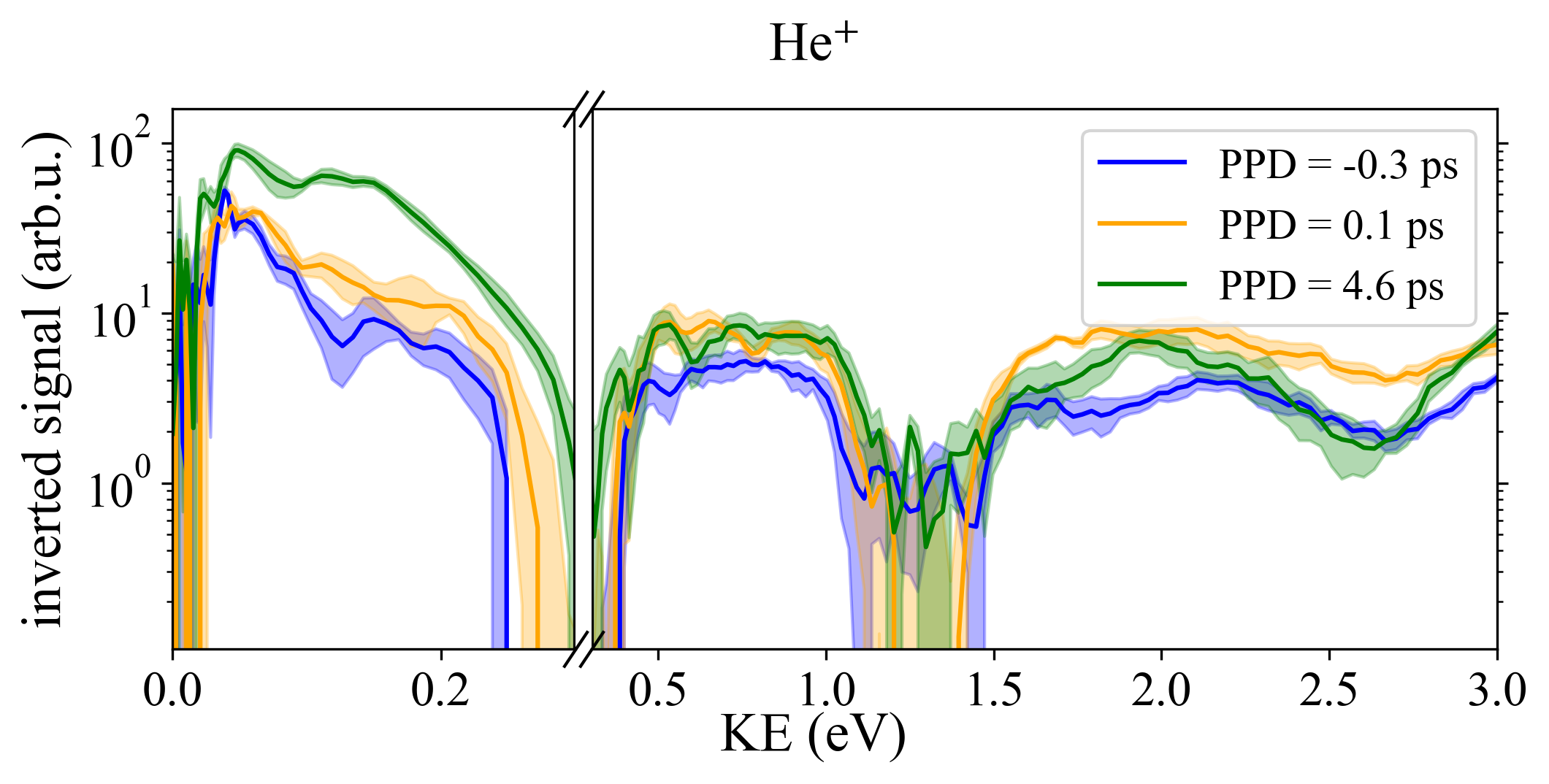}
			\caption[subfig:PlaneGasKER_He+]{\label{subfig:PlaneGasKER_He+_Integrated} }
		\end{subfigure}%
		~
		\begin{subfigure}[t]{0.26\linewidth}
			\centering
			\includegraphics[width=\linewidth*4/4]{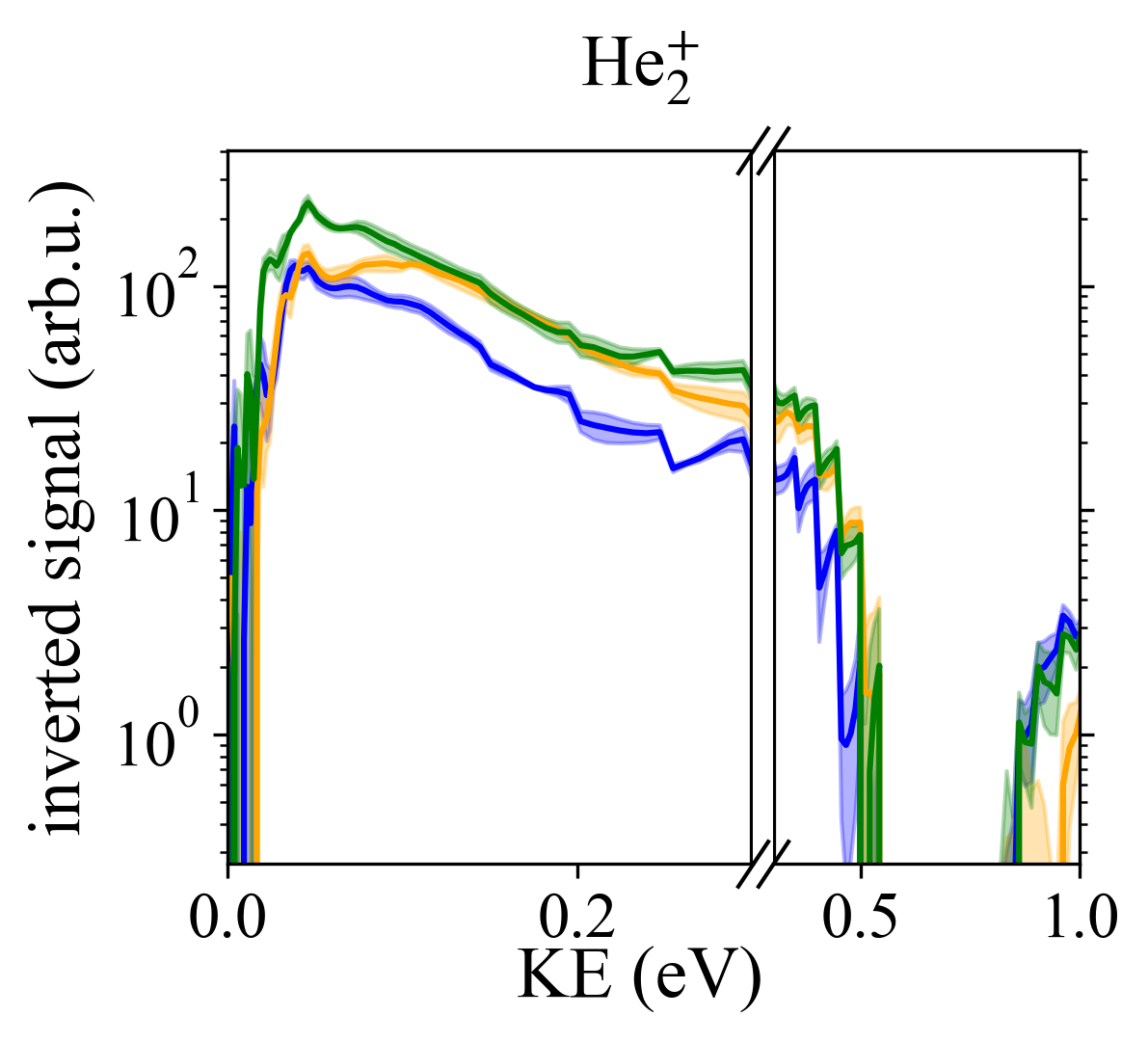}
			\caption[subfig:PlaneGasKER_He2+]{\label{subfig:PlaneGasKER_He2+_Integrated} }
		\end{subfigure}%
		~
		\begin{subfigure}[t]{0.215\linewidth}
			\centering
			\includegraphics[width=\linewidth*4/4]{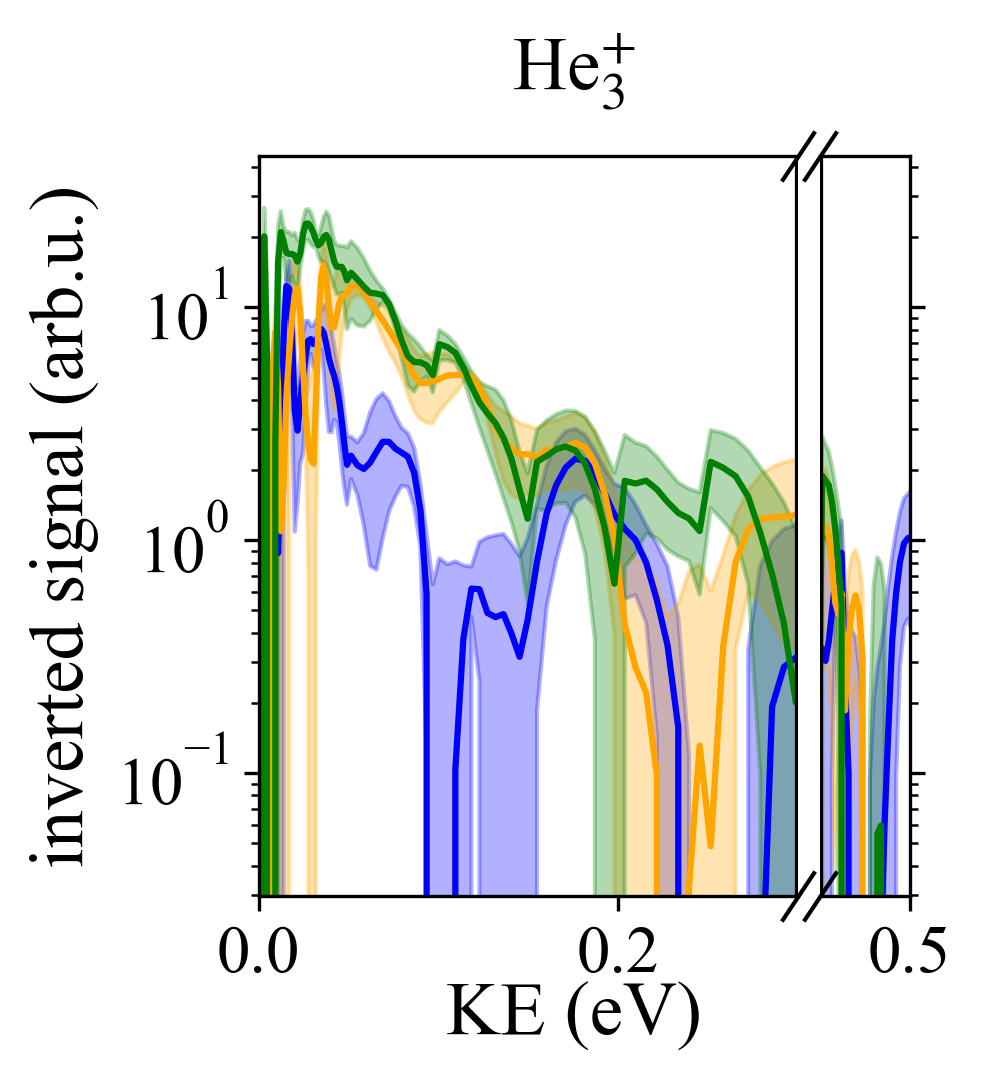}
			\caption[subfig:PlaneGasKER_He3+]{\label{subfig:PlaneGasKER_He3+_Integrated} }
		\end{subfigure}%
	 
		\caption[Tikhonov parameter]{\label{fig:PlaneGasKER} Reconstructed KER spectra for (\subref{subfig:PlaneGasKER_He+}) He$^{+}$ (\subref{subfig:PlaneGasKER_He2+}) He$_{2}^{+}$ (\subref{subfig:PlaneGasKER_He3+}) He$_{3}^{+}$. Black areas denote negative signals (artefacts). Slices of (\subref{subfig:PlaneGasKER_He+_Integrated}) He$^{+}$ (\subref{subfig:PlaneGasKER_He2+_Integrated}) He$_{2}^{+}$ (\subref{subfig:PlaneGasKER_He3+_Integrated}) He$_{3}^{+}$, for a negative, near-zero, and postive pump-probe time-delay (PPD). Shaded regions represent uncertainties in the reconstruction.}
	\end{figure*}

	Through this procedure, the kinetic energy release spectra of the ions were successfully obtained from the TOF spectra, shown as Fig.\ \ref{fig:PlaneGasKER}. 
	The spectra for He$_{2}^{+}$ at KE $>$ 0.8\,eV and He$_{3}^+$ at KE $>$ 0.4\,eV contain artifacts due to the overlap between the He$_{2,3}^+$ peaks in the TOF spectrum. This overlap causes the inversion to produce artifacts at high kinetic energies; the basis functions for a single mass-to-charge ratio are linearly independent, but this may not be true when combining basis functions from two different mass-to-charge ratios. We note that the width of the ion TOF peaks and therefore the overlap of signals from adjacent masses is mostly determined by the extraction time relative to the total time-of-flight. Overlap can be reduced by increasing the extraction voltages, at the cost of reduced ion kinetic energy resolution.

	The resolution of the ion KE spectra is sufficient to distinguish different dynamics of different ion kinetic energy components of the He$^+$ ion signals. The two discrete components at 0.8\,eV and 2.0\,eV arising from the ejection of excited helium atoms from the helium nanodroplets (Fig.\ \ref{fig:PlaneGasKER}\subref{subfig:PlaneGasKER_He+}) could not be studied individually previously. The clear separation of the ion kinetic energy contributions, which allows the unambiguous analysis of the dynamics of the underlying processes, highlights the advantage of our method to retrieve the ion kinetic energy distribution over ad-hoc methods retrieving only average kinetic energy values. For He$_{2}^{+}$ and He$_{3}^{+}$, the presence of low-energy ions $<$ 0.5\,eV can still be attributed to a vibrational excitation of the auto-ionizing He$_{n\geq 2}^{+}$ states \cite{Asmussen2021}.

	\section{Necessity of the reconstruction steps}

	The complexity of the presented reconstruction procedure mainly originates from determining the spatial source distribution $\mu({\bf{x}})$, the spatial-velocity TOF map $T({\bf{x}},{\bf{v}})$, and the instrument function $h(t)$. Once these experimental parameters are known, the determination of basis functions and the inversion process are relatively straightforward. To show that these steps are necessary, we compare other simplifications for $\mu({\bf{x}})$, $T({\bf{x}},{\bf{v}})$, and $h(t)$ with our full simulation:

	\begin{enumerate}
		\setlength\itemsep{0em}
	    \item Assume 1D trajectories for the ions instead of the full 3D simulation of trajectories, so that $T({\bf{x}},{\bf{v}})=T(z,v_z)$ (abbr.\ ``1D");
	    \item Assume a point-like ionization volume instead of considering the widths of the laser and gas jet, so that $\mu({\bf{x}})=\delta^3({\bf{x_0}})$ (abbr.\ ``Point vol.'');
	    \item Assume the measured ion TOF signal is the true TOF signal i.e.\ $h(t)=\delta(t)$ (abbr.\ ``No instrum.'');
	    \item Use roughly known nominal values of experimental parameters instead of determining them through calibration (abbr.\ ``Nom.\ params.'');
	    \item Use full simulation: 3D ion trajectories, expected ionization volume profile, convolution with the instrument function, and using best-fit parameters of the simulation calibrated with the atomic He$^{+}$ TOF spectrum (abbr.\ ``Full sim.'').

	\end{enumerate}

	\begin{figure}
		\begin{subfigure}{\linewidth}
		\centering
		\includegraphics[width=\linewidth*9/10]{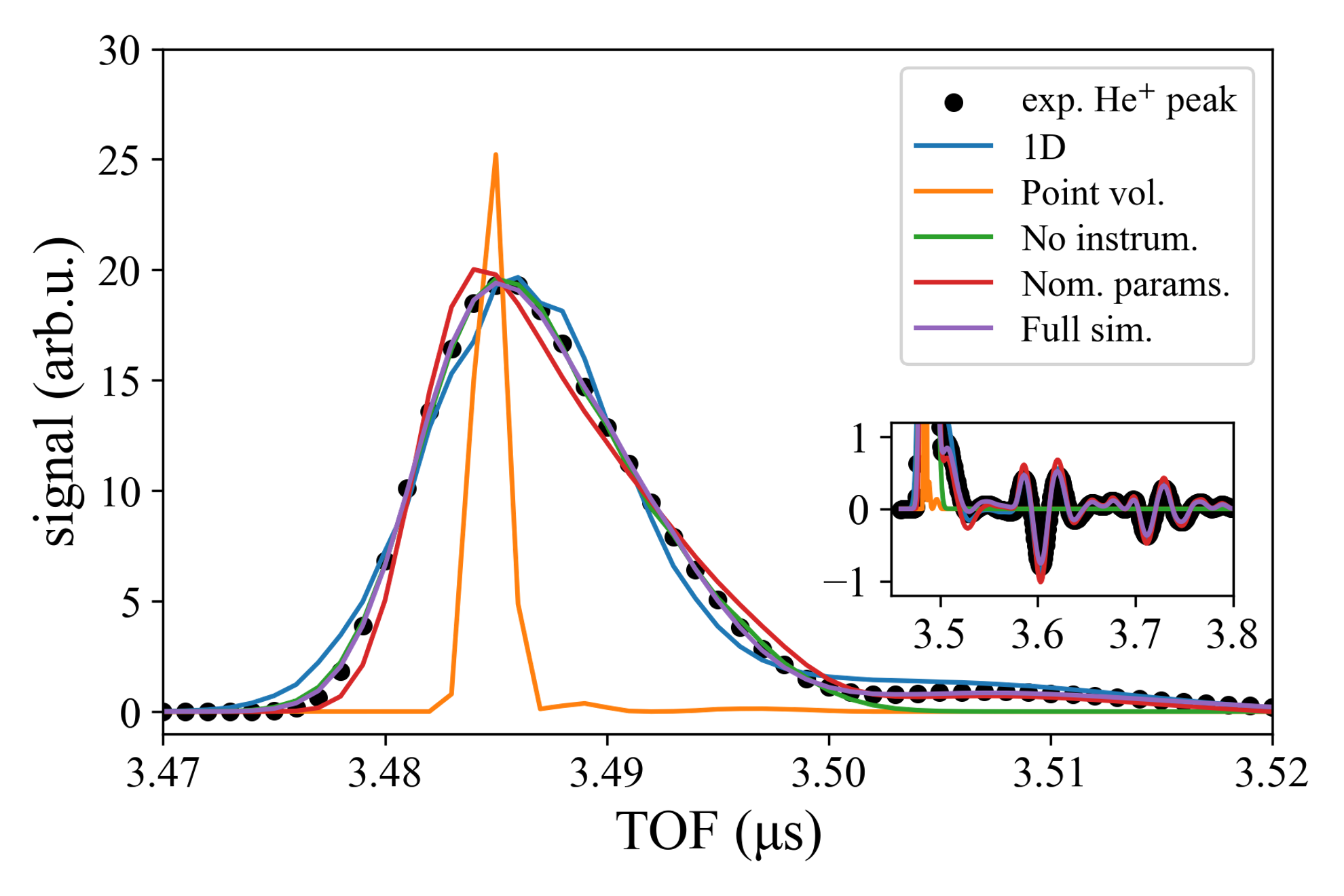}
		\caption[CalibrationComparisons_TOF]{\label{subfig:CalibrationComparisons_TOF}}
		\end{subfigure}%
		
		\begin{subfigure}{\linewidth}
			\centering
			\includegraphics[width=\linewidth*9/10]{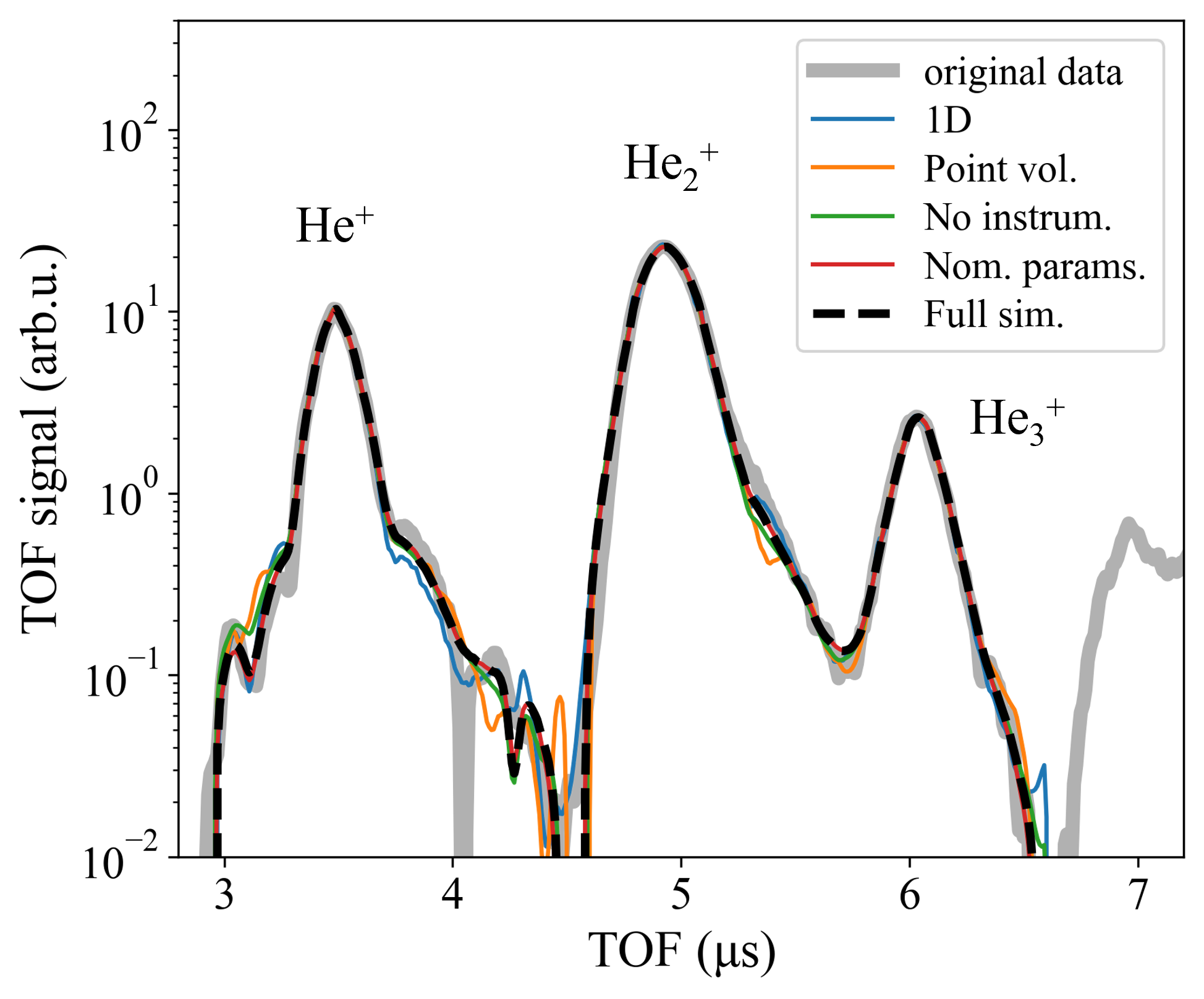}
			\caption[ReconstructionCaseComparison_TOF]{\label{subfig:ReconstructionCaseComparison_TOF} }
		\end{subfigure}%
		
		\begin{subfigure}{\linewidth}
			\centering
			\includegraphics[width=\linewidth*9/10]{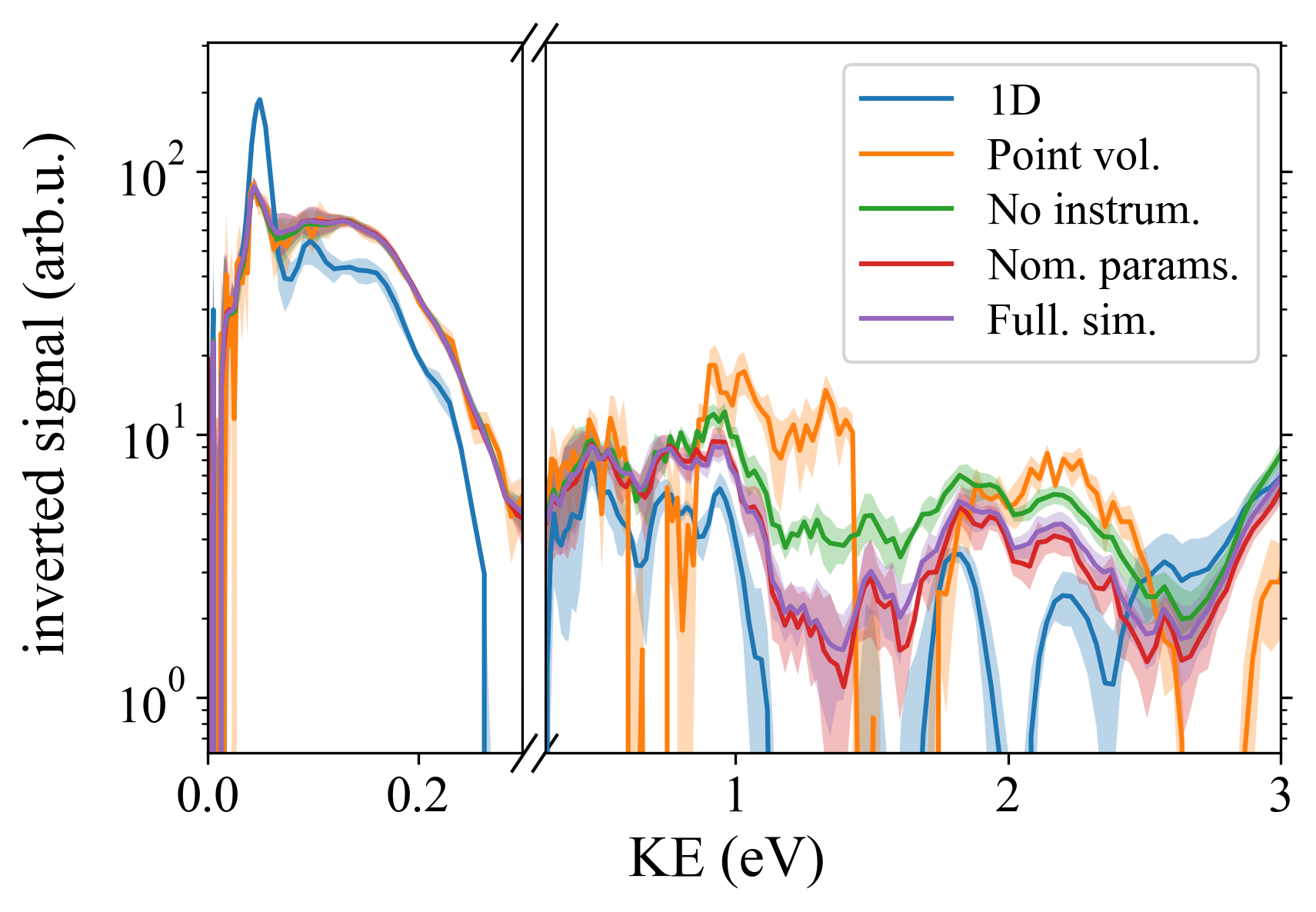}
			\caption[ReconstructionCaseComparison_KER_He+]{\label{subfig:ReconstructionCaseComparison_KER_He+} }
		\end{subfigure}%

	    \caption[CalibrationComparisonsTOF]{\label{ReconstructionComparisons} Comparison of the four simplifications (see text): ``One-dimensional", ``Point volume", ``Nominal parameters", ``No instrument function" with the ``Full simulation'' case in the: (\subref{subfig:CalibrationComparisons_TOF}) reconstruction of the zero-KE calibration peak,  (\subref{subfig:ReconstructionCaseComparison_TOF}) reconstruction of an energy-broadened TOF spectrum, and (\subref{subfig:ReconstructionCaseComparison_KER_He+}) KER reconstruction for He$^{+}$. Black dots correspond to the experimental TOF spectrum. Black dashes correspond to the no-simplification case. The shaded regions represent the uncertainties of the underlying reconstruction.}
		
	\end{figure}

	For each of these simplifying cases, we re-perform the whole reconstruction procedure, with only its respective simplification, in order to represent its best-case scenario. We evaluate the cases by comparing the reconstructed TOF spectra with the original TOF spectrum, and by comparing the reconstructed KER spectra between the cases in Fig.\ \ref{ReconstructionComparisons}.

	We observe that the first two cases, ``1D" and ``Point source" cannot reconstruct parts of the TOF spectrum well. ``Point source'' does not capture the width of the calibration peak in Fig.\ \ref{subfig:CalibrationComparisons_TOF}, which is expected for a non-Wiley-McLaren geometry. Although the ``1D'' case seems plausible there, it fails to correctly reproduce the shape of the kinetic-energy-broadened TOF spectrum in Fig.\ \ref{subfig:ReconstructionCaseComparison_TOF}. The consequence on the KER spectra is immediately clear; they are noticeably dissimilar. This implies for the trajectories that the motion perpendicular to the spectrometer axis, as well as the size of the ionization volume, has a significant effect on the shape of the TOF spectrum.

	The third case, ``no instrument function" reproduces the leading edges of the original TOF spectrum well, with noticeable errors only occurring away from the peak centres. In the KER spectrum, this manifests as similarly small errors in the high kinetic energy range where the reconstructed signal is low, compared to the ``Full simulation" case.

	The most significant change between the fourth ``Nominal parameters" case and the ``Full simulation" case is a smaller source volume $\mu({\bf{x}})$ (30 $\upmu$m vs. 45 $\upmu$m), determined by the laser width. Despite this difference, the reconstructed TOF and KER spectra are nearly identical to one another, which shows the stability of the reconstruction with respect to small changes in the experimental parameters. e.g.\ spatial broadening.

	To further show that the reconstruction is stable with respect to a change in the 
	shape of the ionization volume, we provide an additional analysis with a different shape for the ionization volume in Appendix D in the Supplementary Material, which yields nearly identical results.

	\paragraph*{Summary}

	We demonstrated a method to reconstruct KER spectra from one-dimensional TOF spectra, and show how different simplifications to this procedure may lead to erroneous reconstructions. With our spectrometer geometry, we found experimental parameters through a calibration procedure involving ion trajectory simulations, and we extrapolated these trajectories to form basis functions in the TOF coordinate, indexed by initial velocities. These basis functions were used to create an inversion matrix, which we applied to TOF spectra from a previous experiment. These KER reconstructions reproduce an additional physical feature with a similar timescale to a relaxation to the droplet 1s2s ${}^{3}$S electronic state seen in Ref. \cite{Asmussen2021}, showing that quantitative characterization of kinetic-energy features from TOF spectra are feasible within certain constraints, yielding more information than the usual previous treatments of mass-to-charge ratio characterization. We further note that this specific spectrometer has been used in numerous other experiments e.g.\ in Refs.\ \cite{Squibb2018, Laforge2019}, and this technique could be used to supplement previously published as well as future results, through the analysis of ion TOF data, to yield ion KER spectra.


 \section*{Supplementary Material}

 The supplementary material contains examples of different regularization parameters, a comment on the TOF map $T(r,z)$ for a cylindrically-symmetric potential, discussion on the uniqueness of the determination of parameters from the calibration procedure, and fit parameters of a different assumed shape of the ionization volume.

\begin{acknowledgments}
	We gratefully acknowledge funding from the Deutsche Forschungsgemeinschaft (grant numbers STI 125/19-2, RTG 2717) and the COST Action CA21101 “Confined Molecular Systems: From a New Generation of Materials to the Stars (COSY)”. We additionally acknowledge the participants of the experiment performed in Ref. \cite{Asmussen2021}.
\end{acknowledgments}

\section*{Author Declarations}

\subsection*{Conflict of Interest}

    The authors have no conflicts to disclose.
    
\subsection*{Data Availability}
 
    Raw data were generated at FERMI at the Elettra Syncrotron large scale facility. Derived data supporting the findings of this study are available from the corresponding author upon reasonable request.

\bibliography{TOF_InversionMethodologyReferences}

\end{document}


\preprint{AIP/123-QED}

\title[Supplementary Material: Kinetic energy reconstruction from time-of-flight spectra]{Supplementary Material for Method of Kinetic energy reconstruction from time-of-flight spectra}
\author{A. Ngai}
\affiliation{Institute of Physics, University of Freiburg, Hermann-Herder-Str. 3, 79104 Freiburg, Germany}
\author{K. Dulitz}
\affiliation{Institut für Ionenphysik und Angewandte Physik, Universität Innsbruck, 6020 Innsbruck, Austria}
\author{S. Hartweg}
 \email{Sebastian.Hartweg@physik.uni-freiburg.de.}
\affiliation{Institute of Physics, University of Freiburg, Hermann-Herder-Str. 3, 79104 Freiburg, Germany}
\author{J. Franz}
\affiliation{Institut für Physik, Universität Kassel, Heinrich-Plett-Straße 40, 34132 Kassel, Germany}
\author{M. Mudrich}
\affiliation{Department of Physics and Astronomy, Aarhus University, Ny Munkegade 120, 8000 Aarhus C, Denmark}
\author{F. Stienkemeier}
\affiliation{Institute of Physics, University of Freiburg, Hermann-Herder-Str. 3, 79104 Freiburg, Germany}

\date{\today}
             
\maketitle

\appendix

    \section{Choice of the regularization parameter}
    \label{appendix:Regularization}
		
\begin{figure}[t]
			\centering
			\begin{subfigure}[t]{0.5\linewidth}
				\centering
				\includegraphics[width=\linewidth*4/4]{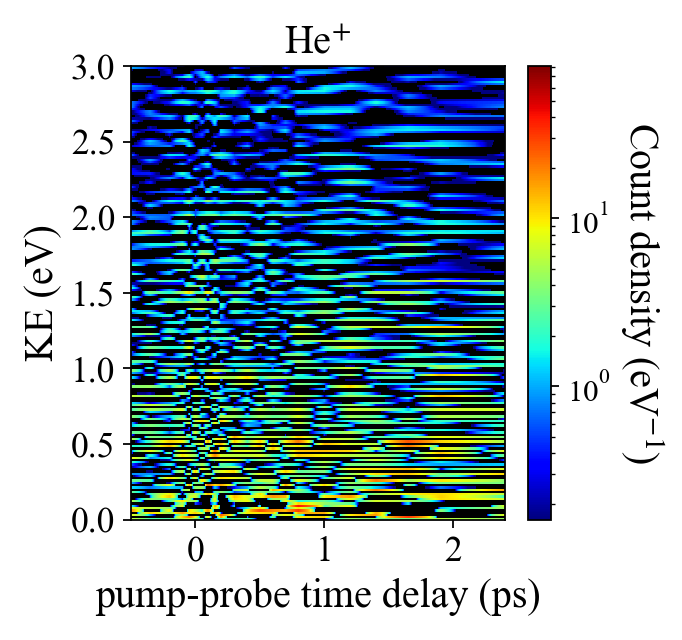}
				\caption[LoTikhonovHe+_placeholder]{\label{LoTikhonovHe+} }
			\end{subfigure}%
			~
			\begin{subfigure}[t]{0.5\linewidth}
				\centering
				\includegraphics[width=\linewidth*4/4]{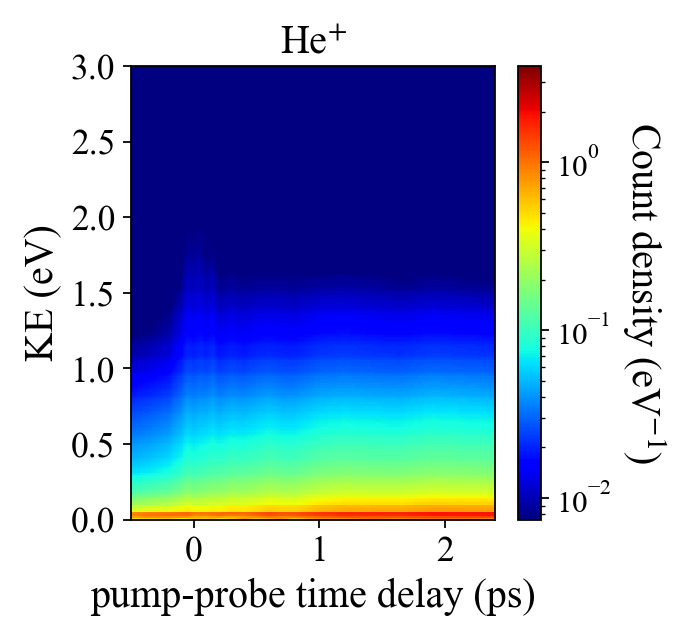}
				\caption[HiTikhonovHe+]{\label{HiTikhonovHe+} }
			\end{subfigure}%
		
			\begin{subfigure}[t]{0.5\linewidth}
				\centering
				\includegraphics[width=\linewidth*4/4]{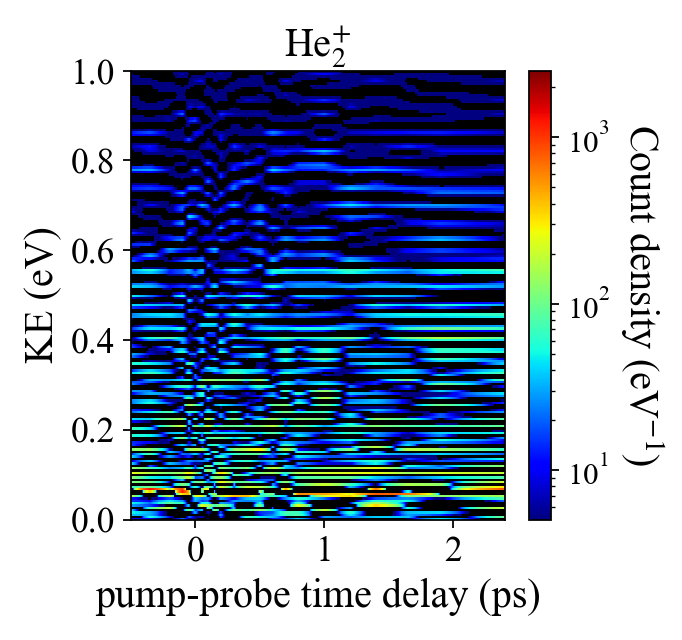}
				\caption[LoTikhonovHe2+_placeholder]{\label{LoTikhonovHe2+} }
			\end{subfigure}%
			~
			\begin{subfigure}[t]{0.5\linewidth}
				\centering
				\includegraphics[width=\linewidth*4/4]{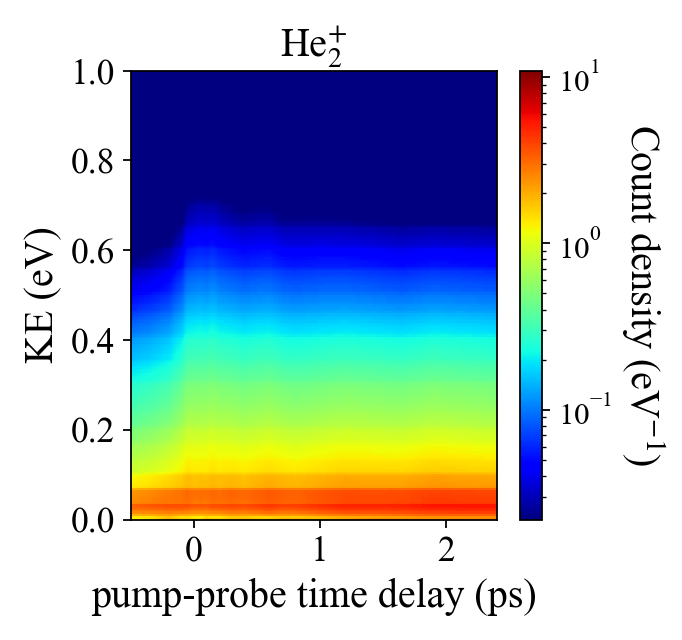}
				\caption[HiTikhonovHe2+]{\label{HiTikhonovHe2+} }
			\end{subfigure}%
		
			\begin{subfigure}[t]{0.5\linewidth}
				\centering
				\includegraphics[width=\linewidth*4/4]{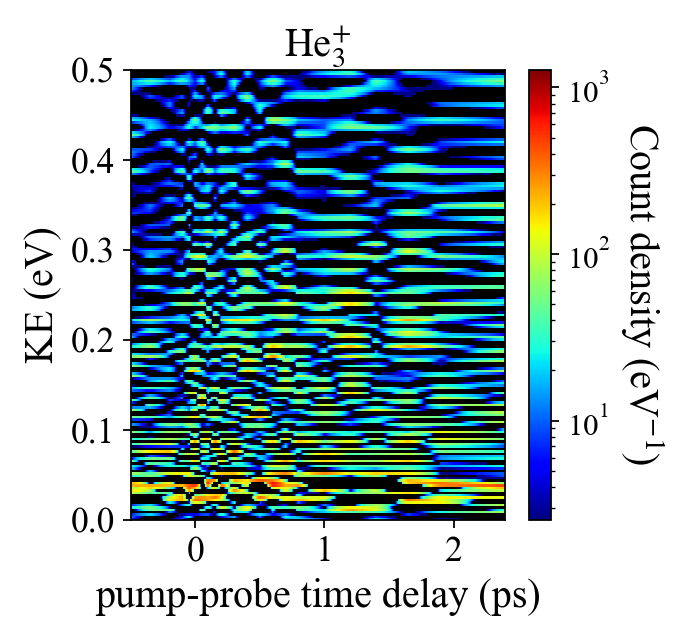}
				\caption[LoTikhonovHe3+_placeholder]{\label{LoTikhonovHe3+} }
			\end{subfigure}%
			~
			\begin{subfigure}[t]{0.5\linewidth}
				\centering
				\includegraphics[width=\linewidth*4/4]{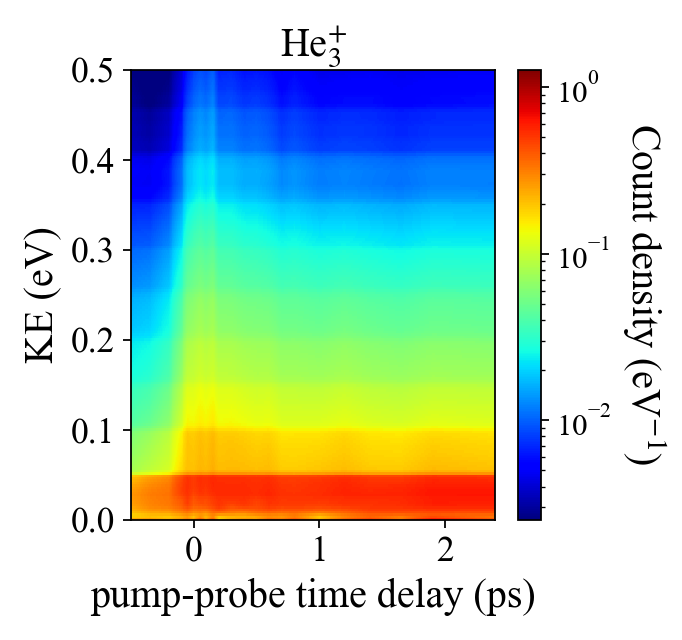}
				\caption[HiTikhonovHe3+]{\label{HiTikhonovHe3+} }
			\end{subfigure}%
		
			\caption[Tikhonov parameter]{\label{VaryingRegularization} Under-regularization for (\subref{LoTikhonovHe+}) He$^{+}$ (\subref{LoTikhonovHe2+}) He$_{2}^{+}$ (\subref{LoTikhonovHe3+}) He$_{3}^{+}$.  Over-regularization for (\subref{HiTikhonovHe+}) He$^{+}$ (\subref{HiTikhonovHe2+}) He$_{2}^{+}$ (\subref{HiTikhonovHe3+}) He$_{3}^{+}$. Compare to the regularization presented in Fig.~(6).}
		\end{figure}

		Ideally, regularization would not be used, as such a scheme biases any solution. Given the certainty of a bias, we must impose conditions on an acceptable solution, and this would bound our regularization parameter.
		With our optimization problem defined as:
		\begin{equation}
	    \tag{11 revisited}
			{\bf{g}} \coloneqq \argmin_{{\bf{g}}} \left\{\left\|{\bf{f}}-{\bf{K}}{\bf{g}}\right\|_2 + \Lambda_0\left\|{\bf{g}}\right\|_2 + \Lambda_1\left\|{\bf{D}}{\bf{g}}\right\|_2\right\} ,
		\end{equation}
		\begin{equation}
	    \tag{12 revisited}
			{\bf{g}}={\bf{T}}{\bf{f}}={\bf{K}}^T \left({\bf{K}}{\bf{K}}^T -\Lambda_0 {\bf{I}}-\Lambda_1 {\bf{D}}^2 \right)^{-1} {\bf{f}} ,
		\end{equation}
	    we define our ``optimal'' regularization parameters ${\Lambda_0,\Lambda_1}$ as the smallest such parameters which satisfy the following heuristic conditions:
		\begin{enumerate}
			\setlength\itemsep{0em}
			\item Features that are above the noise must not be lost 
			\item The KER spectra of an ion must decay at ``large'' KE 
		\end{enumerate}
		The effects of disregarding these two conditions is shown in Fig.\ \ref{VaryingRegularization}, as either using under- or over-regularization.

		\begin{figure}[ht]
			\centering
			\begin{subfigure}[t]{0.45\linewidth}
				\centering
				\includegraphics[width=\linewidth*4/4]{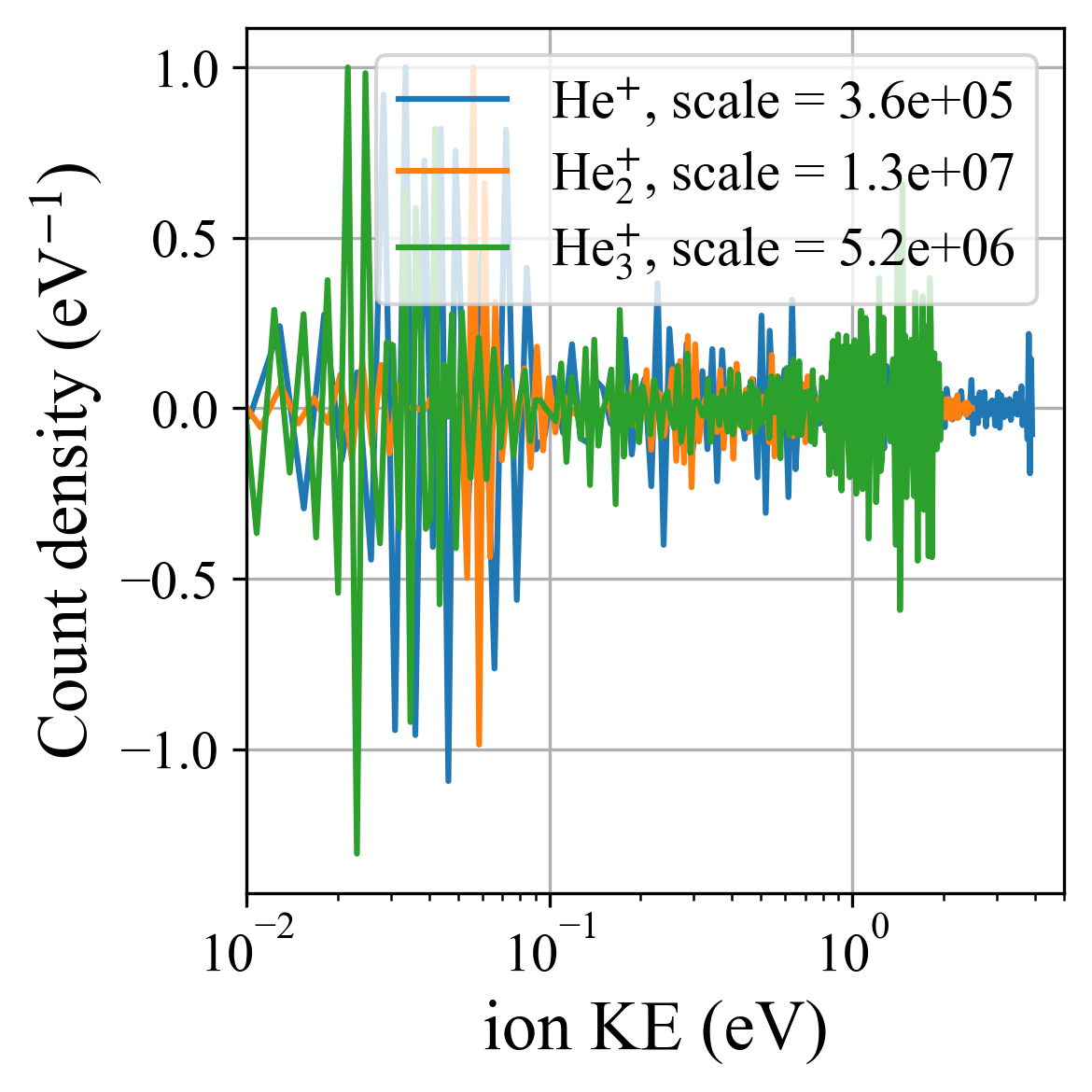}
				\caption[LoTikhonov_placeholder]{\label{LoTikhonovKE} }
			\end{subfigure}%
		~
			\begin{subfigure}[t]{0.45\linewidth}
				\centering
				\includegraphics[width=\linewidth*4/4]{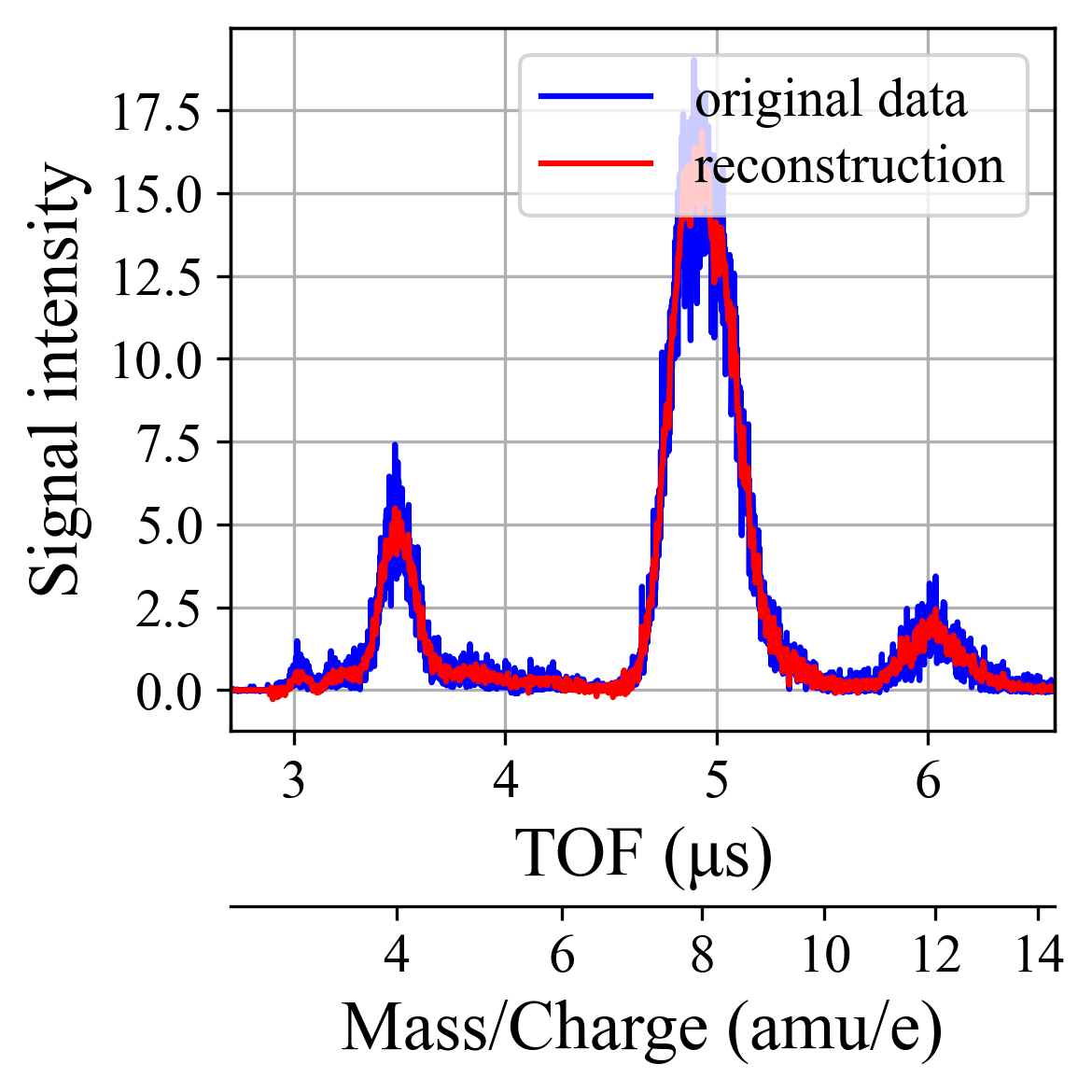}
				\caption[LoTikhonov_placeholder]{\label{LoTikhonovTOF} }
			\end{subfigure}%
			
			\begin{subfigure}[t]{0.45\linewidth}
				\centering
				\includegraphics[width=\linewidth*4/4]{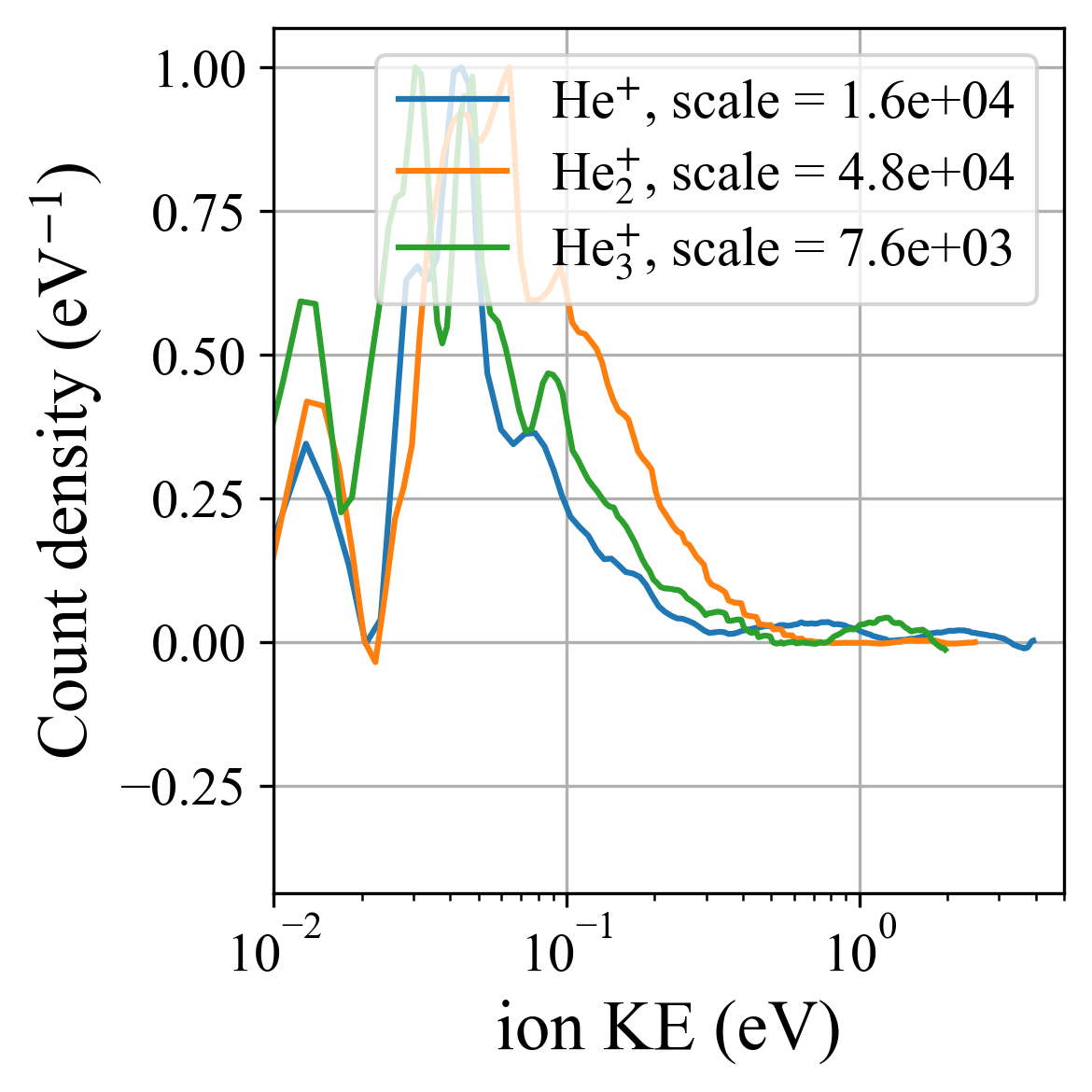}
				\caption[MiTikhonov_placeholder]{\label{MiTikhonovKE} }
			\end{subfigure}%
		~
			\begin{subfigure}[t]{0.45\linewidth}
				\centering
				\includegraphics[width=\linewidth*4/4]{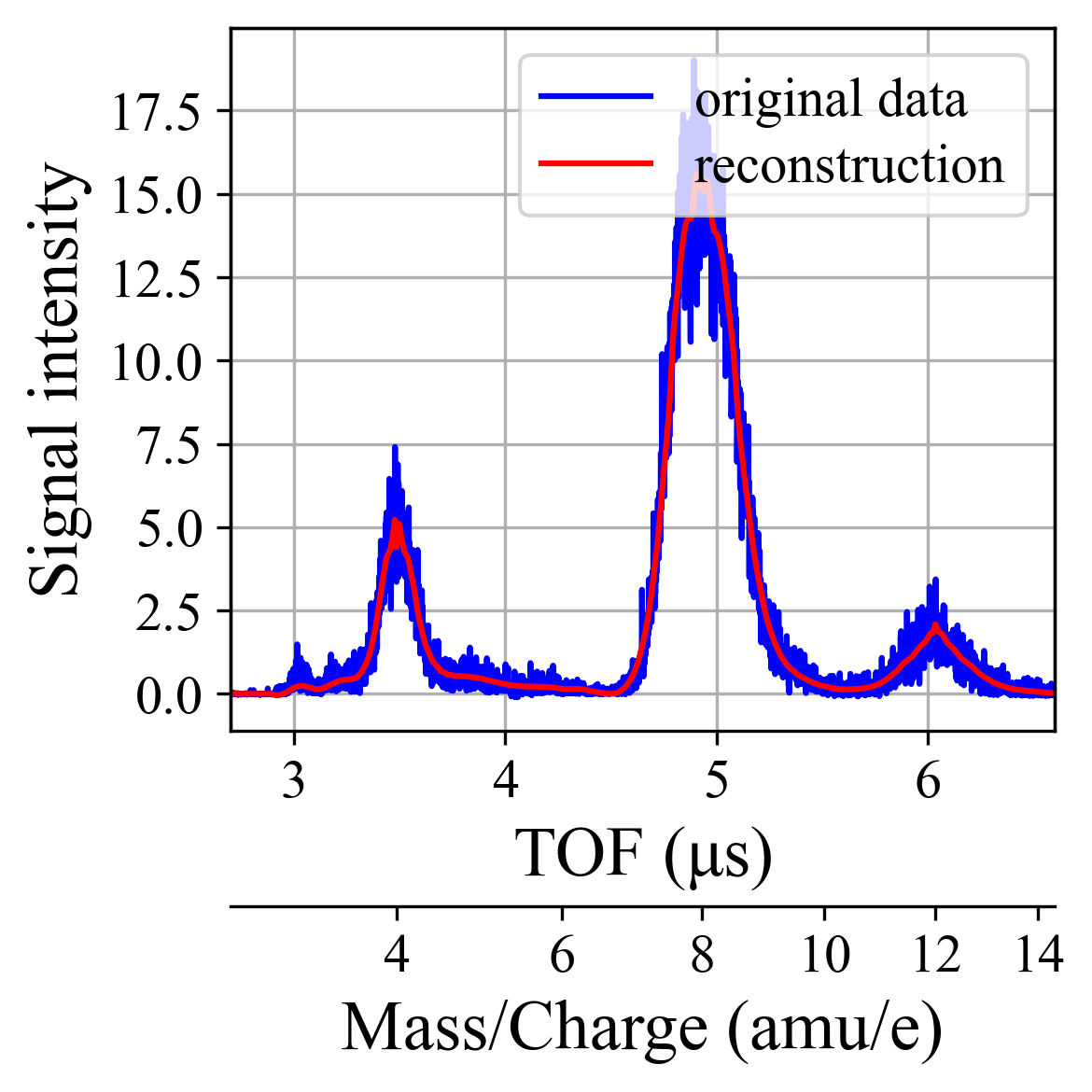}
				\caption[MiTikhonov_placeholder]{\label{MiTikhonovTOF} }
			\end{subfigure}%
			
			\begin{subfigure}[t]{0.45\linewidth}
				\centering
				\includegraphics[width=\linewidth*4/4]{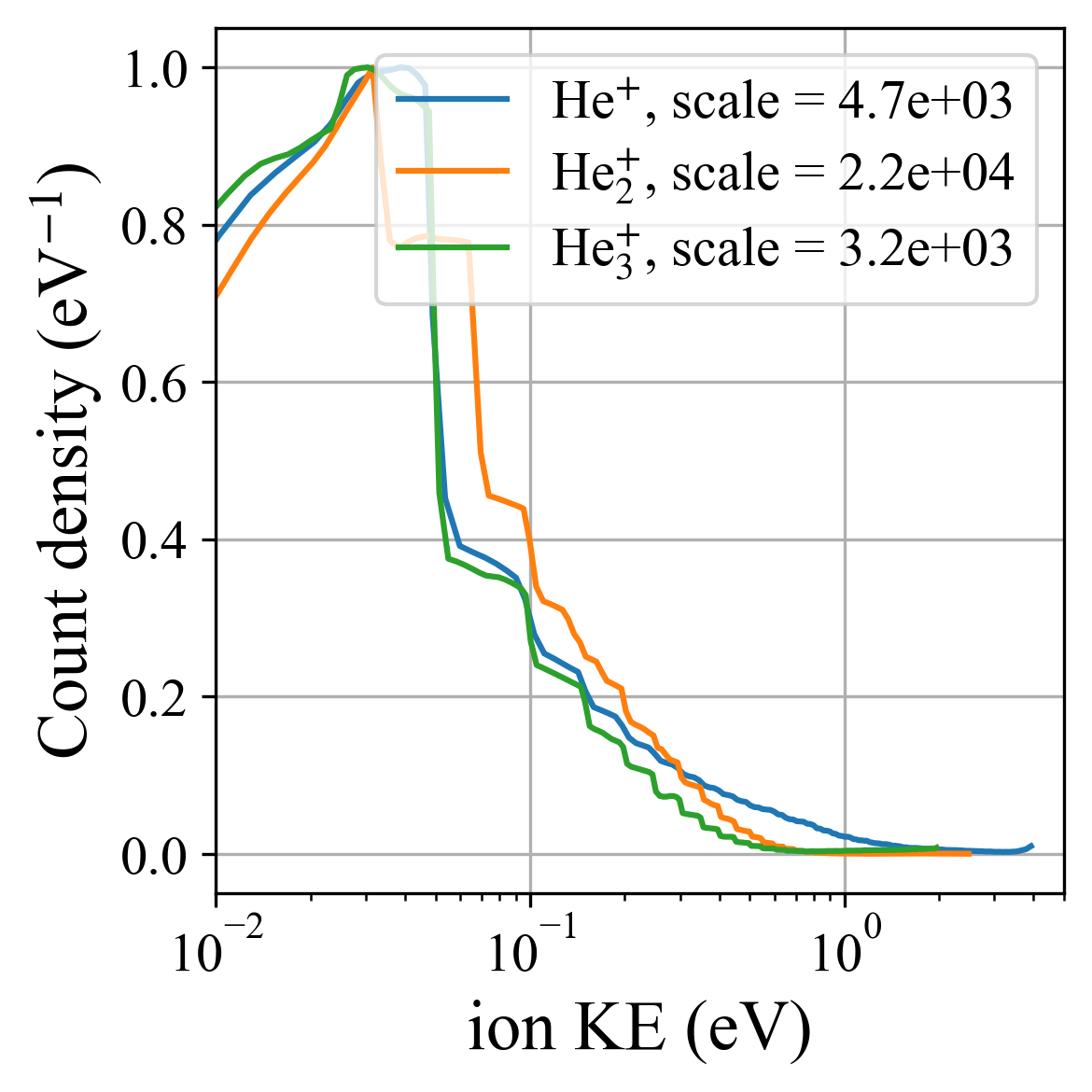}
				\caption[HiTikhonov]{\label{HiTikhonovKE} }
			\end{subfigure}%
		~
			\begin{subfigure}[t]{0.45\linewidth}
				\centering
				\includegraphics[width=\linewidth*4/4]{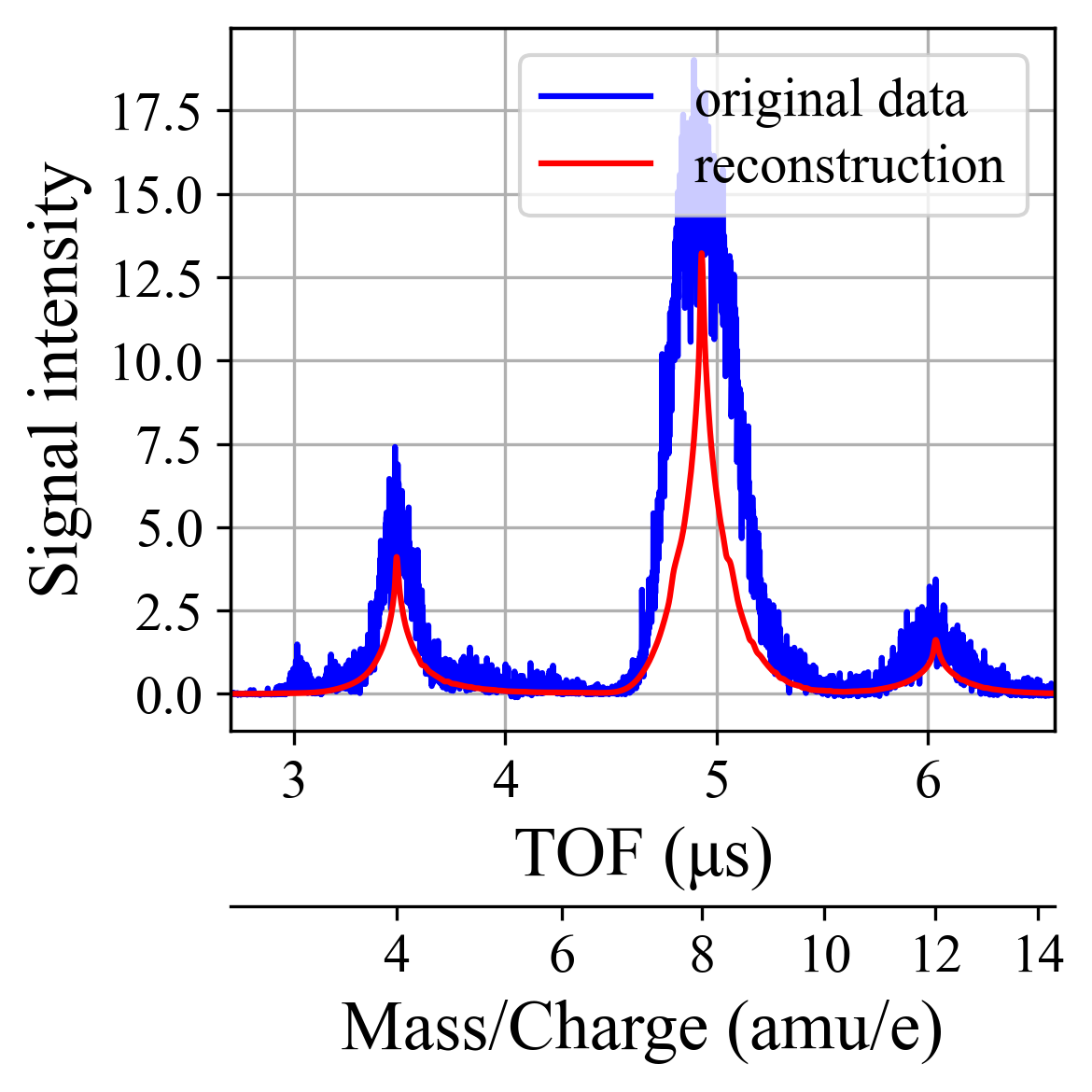}
				\caption[HiTikhonov]{\label{HiTikhonovTOF} }
			\end{subfigure}%
		\caption[Tikhonov parameter]{\label{fig:VaryingRegularizationTraces} Slice of Fig.\ \ref{VaryingRegularization} at pump-probe delay=0.5 ps, showing the effect of the regularization parameter on the reconstructed TOF trace. (\subref{LoTikhonovKE})(\subref{LoTikhonovTOF}) shows the reconstructed KE and TOF spectra without regularization, (\subref{MiTikhonovKE})(\subref{MiTikhonovTOF}) with acceptable regularization, (\subref{HiTikhonovKE})(\subref{HiTikhonovTOF}) with too large regularization. }
		\end{figure}

		\section{Approximation for $T(r,z)$ within a cylindrically-symmetric potential}
	    \label{appendix:LaplaceSymmetric}
		
		The idea behind Eq.~(17) is that the spatial TOF map for zero ion KE $T(r,z)$ is well-approximated by a Taylor expansion around $(r=0,z=z_0)$. The potential $\Phi({\bf{x}})$ is found through the Laplace equation $\nabla^2 V=0$, of which the case of cylindrical symmetry and boundary conditions on a cylindrical volume can be solved numerically (see e.g. Ref. \cite{Heerens1982}). Having $\Phi(r,x)$, the TOF at different initial starting positions can be calculated.  The coefficients in Eq.~(17) are obtained as:
		\begin{equation}
        \label{eq:TOF_Taylor}
			T(r,z)=\sum_{m,n=0}^\infty \partial_r^m \partial_z^n T(r,z)\bigg|_{(r=0,z=z_0)} r^m z^n .
		\end{equation}
		To simplify matters, let us make an additional assumption about the trajectories: ${\bf{x}}(t)=(r_0,z_0+f(t))$ i.e.\ we assume that particles only fly along the z-axis, which is approximately satisfied by potential fields with only a slight radial curvature. This simplification allows us to directly calculate the coefficients in Eq.~\eqref{eq:TOF_Taylor} from $\Phi(r,x)$ using Eq.~(3).
  
        Because of the cylindrical boundary condition $\partial_r \Phi(r=0,z)=0$, all terms $A_{m,n}\coloneqq \partial_r^m \partial_z^n T(r,z)\bigg|_{(r=0,z=z_0)} $ with odd $m$ are zero, and we arrive at Eq.~(17).

		\begin{figure}
			\centering
		\includegraphics[width=\linewidth*3/4]{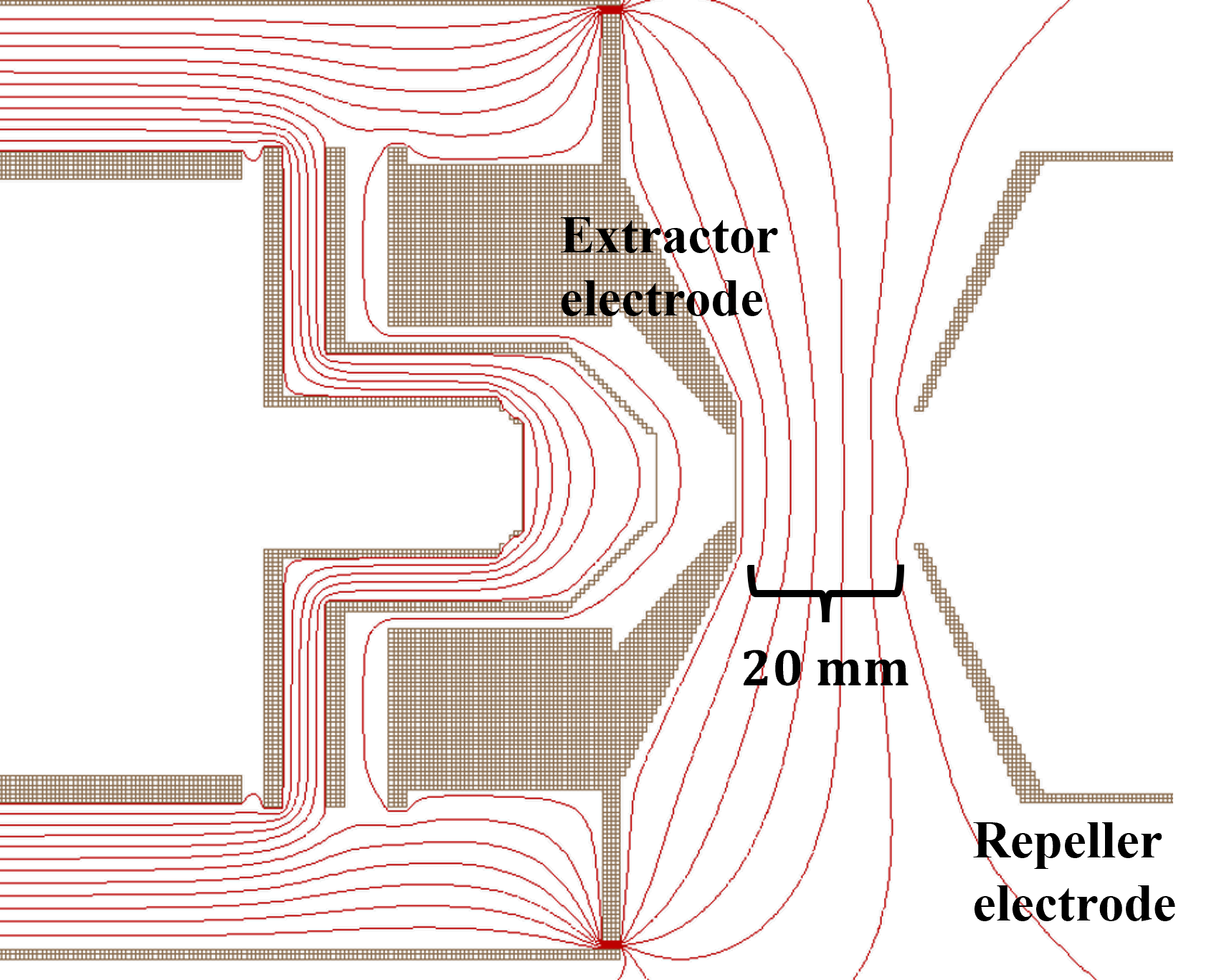}
		\caption[fig:SymmetrySchematic]{\label{fig:SymmetrySchematic} 2D slice along the symmetric axis of the electrode geometry in SIMION, used for our ion trajectory simulations.}
		\end{figure}

    \section{Uniqueness of the calibration procedure}
	    \label{appendix:CalibrationUncertainties}
     
	    \subsection{Uniqueness Conditions for Calibration of Experimental Parameters}
        \label{appendix:CalibrationUniquenessConditions}
	    
	    In the calibration procedure, we use test spectra to match (calibrate) adjustable parameters in our trajectory simulations to the apparent experimental conditions. For these simulations to be reliable, our calibration procedure must yield a unique result.

        To show this, let us consider a set of adjustable experimental parameters $P$ with corresponding directional derivatives $\partial P$:
        \begin{equation}
        \begin{aligned}
		P &\coloneqq \{P_1,...,P_{m+n}\} ,
        \\
		\partial P &\coloneqq \{\partial_{P_1},...,\partial_{P_{m+n}}\} ,
        \end{aligned}
        \end{equation}
	    and rewrite Eq.~(4) as:
	    \begin{equation}
	    \label{LambdaDefinition}
		f(t)=\iint d^3{\bf{v}}\ d^3{\bf{x}}\ \Lambda({\bf{x}},{\bf{v}}; P) .
	    \end{equation}
	    We now formulate necessary conditions for Eq.~\eqref{LambdaDefinition} to guarantee uniqueness for our calibration procedure:
	    \begin{itemize}
		\setlength\itemsep{0em}
		\item Condition 1 (Fredholm invertibility): $\Lambda \leftrightarrow f$ is bijective for a given $P$,
		\item Condition 2 (Commutativity): the elements in $\partial P$ commute,
		\item Condition 3 (Linear-independence): the actions of $\partial P$ on $\Lambda$ are all linearly-independent.
	    \end{itemize}
	    Condition 1 allows conditions 2 and 3 to imply the existence of a coordinate map through the following corollary (contraposition to Prop 8.11b, Theorem 9.46 in Ref.\ \cite{Lee2012}):
		\begin{corollary} Let $f_{p}: \mathbb{R}\rightarrow \mathbb{R}$, and $p\in M$ where $M$ is a simply-connected $N$-dimensional differentiable manifold. $f_p$ is uniquely parameterized by $p$ if and only if there exists a set of basis vectors $\partial P\coloneqq \{\partial_{P_i}\in T_p M\ |\ i=1,...,N \}$ for the tangent space $T_{p}M$ on $M$ over point $p$ such that:
		\begin{enumerate}
		    \item $[\partial_{P_i},\partial_{P_j}]=0\ \forall\  \partial_{P_i},\partial_{P_j}\in \partial P$ \\ (commutativity $\Leftrightarrow$ trivial Lie bracket)
		    \item $\dim(\mathspan(\partial P)) = \dim(M)$ \\ (linear-independence)
		\end{enumerate}
		\label{CoordinateTheorem}
		\end{corollary}
	    In other words, these conditions allow us to use the parameters $P$ as a coordinate chart that uniquely maps parameters to TOF spectra $P\leftrightarrow f(t)$. These conditions can be satisfied as follows:
	
	    \textit{Condition 1:} Eq.~\eqref{LambdaDefinition} is known as a Fredholm integral of the first kind and is not invertible in general. We instead consider an approximate problem which is invertible via substituting our direct inversion problem with a discretized least-squares inversion problem, e.g.\ with the Moore-Penrose pseudoinverse (Eq.~(10)), which can be more reliably constructed than a direct inverse, or with a regularization scheme (Eq.~(12)) whose invertibility is guaranteed for a large-enough regularization parameter.
	    
	    \textit{Condition 2:} Assuming the second derivative of $\Lambda$ with respect to $P$ is continuous, then by Schwarz's theorem the partial derivatives commute. This is the case for ionization volumes located entirely within the confines of a "well-behaved" spectrometer.
	    
	    \textit{Condition 3:} This must be checked for every different experimental geometry. Should this condition not be fulfilled, linear independence can be restored through the following: for every redundant degree of freedom, there must be an additional constraint which is independent of this calibration procedure.

        In practice, only condition 3 must be checked. An example is given in the next subsection.

	    \subsection{Linear independence in the gas jet/laser crossed-beam experimental geometry}
	    \label{appendix:QuantityDerivation}
     
	    
	    \paragraph{Checking linear independence of $\partial P$}
	    
	    As the TOF spectra come from Eq.~(4), we consider a set of (adjustable experimental) parameters $P_{\mu} = \{\mu_1, ...,\mu_m\}$ and $P_{\nu} = \{\nu_1, ...,\nu_n \}$:
	    \begin{enumerate}
		\setlength\itemsep{0em}
		\item $P_{\mu}$ characterizes the ionization volume (gas jet and laser spatial profile).
		\item $P_{\nu}$ characterizes the voltage fields.
	    \end{enumerate}
	    They affect the initial spatial distribution $\mu({\bf{x}}; P_{\mu} )$ and the TOF $T({\bf{x}}; P_{\nu})$ respectively. For simplicity, we will also address them simultaneously as the set $P=P_{\mu} \cup P_{\nu}$.

	    In order to show that $P$ maps onto unique TOF spectra, we look at Eq.~(4) again. If we assume cylindrical symmetry, the integral is simplifies to:
	    \begin{equation}
	    \label{TOF_TransformationCylindricalSummed}
		f(t) = 2\pi \int_0^\infty \!\! dr \left\{r \sum_{z:t=T}\mu(r,z)\frac{\sqrt{1+\left(\frac{dz}{dr}\right)^2}} {\sqrt{(\frac{\partial T}{\partial z})^2 + (\frac{\partial T}{\partial r})^2}}\right\}.
	    \end{equation}
	    %
	    For simplicity, we require that in the region where $\mu(r,z)\neq 0$: for a fixed $r$, $T(r,z)=t$ only has one solution for $z$, i.e.\ is single-valued. We then simplify Eq.~\eqref{TOF_TransformationCylindricalSummed} as Eq.~(18):
	    \begin{equation}
	    \tag{18 revisited}
		\begin{aligned}
		f(t) &=2\pi \int_0^\infty \!\! dr \left\{r\ \mu(r,z)\left|\frac{\partial T(r,z)}{\partial z }\right|^{-1}\right\}_{(z(r,t):t=T)}.
		\end{aligned}
	    \end{equation}
	    Note that the integration could also be performed on the $r$- instead of the $z$-coordinate.  We define the integrand as a function $\Lambda=\Lambda(r;P)$:
	    \begin{equation}
	    \label{LambdaIntegrandDefinition}
		\Lambda(r;P) \coloneqq  r\ \mu\left(r,z;P_{\mu}\right)    \left|\frac{\partial T(r,z;P_{\nu})}{\partial z }\right|^{-1} \bigg|_{(z(r,t):t=T)},
	    \end{equation}
	    
        According to Appendix \ref{appendix:CalibrationUniquenessConditions}, we only have to show linear-independence of the partial derivatives. This linear-independence condition is written as:
	    \begin{equation}
		\sum_i c_i \partial_i \Lambda=0 \Rightarrow c_i = 0 ,
	    \end{equation}
	    where $\Lambda$ depends on $\mu(r,z;P_{\mu})$ and $T(r,z;P_{\nu})$. The function $\mu(r,z;P_{\mu})$ depends on the experimental geometry, while $T(r,z;P_{\nu})$ can be discussed in general.
	    
	    \paragraph{General elements in $\partial P_{\nu}$}
	    
	    $T(r,z)$ can be approximated as a Taylor expansion:
	    \begin{equation}
	    \tag{17 revisited}
		\begin{aligned}
		T(r,z)&\approx A_{00}+A_{01}z+A_{10}r^2+A_{11}zr^2
		\end{aligned} ,
	    \end{equation}
	    where every term in ${\partial_{A_{mn}} \Lambda :m,n\in \mathbb{N}_0 }$ is linearly independent. It remains to show that the parameters between $T(r,z)$ and $\mu(r,z)$ are also linearly-independent for the specific experimental geometry.
	    
	    \paragraph{Specific elements in $\partial P_{\mu}$: crossed-beam experiment}
	    
	    We will prove that the linear-independence condition holds for a gas jet/laser crossed-beam experiment (Fig.\ (2) gives an example for such a setup), where $\mu(r,z)$ has the following form:
	\begin{equation}
    \tag{16 revisited}
    \begin{aligned}
		\mu({\bf{x}})=\mu(r,z)=I_L(r,z,\theta) I_G(r,z,\theta) ,
        \\
		I_L(x,y,z) = \frac{1}{\sigma_L^2 2\pi} e^{-\frac{1}{2}(\frac{z-z_L}{\sigma_L})^2} e^{-\frac{1}{2}(\frac{x-x_L}{\sigma_L})^2},
        \\
		I_G(x,y,z) = \frac{1}{2 \sigma_G} H\left(\sigma_G - |y-y_G|\right),
    \end{aligned}
	\end{equation}
	    where $H(x)$ is the Heaviside step function, $I_L$ is a Gaussian laser profile along the $y$-axis, and $I_G$ models a planar gas jet profile along the $x,y$-axes (Fig.~2). We then have the following set of overlap parameters: $P_{\mu}=\{x_L,z_L, \sigma_L,y_G,\sigma_G \}$. We also consider the set of potential parameters up to first order in $z$: $P_{\nu}=\{A_{00},A_{10},A_{01},A_{11}\}$. We will show that the only condition necessary for linear independence within the full set of experimental variables  $P=\{x_L,z_L, \sigma_L,y_G,\sigma_G ,A_{00},A_{10},A_{01},A_{11}\}$ is that $z_L$ must be known, or at least constrained.
	    
	\begin{figure}[t]
		\centering
		\includegraphics[width=\linewidth*3/4]{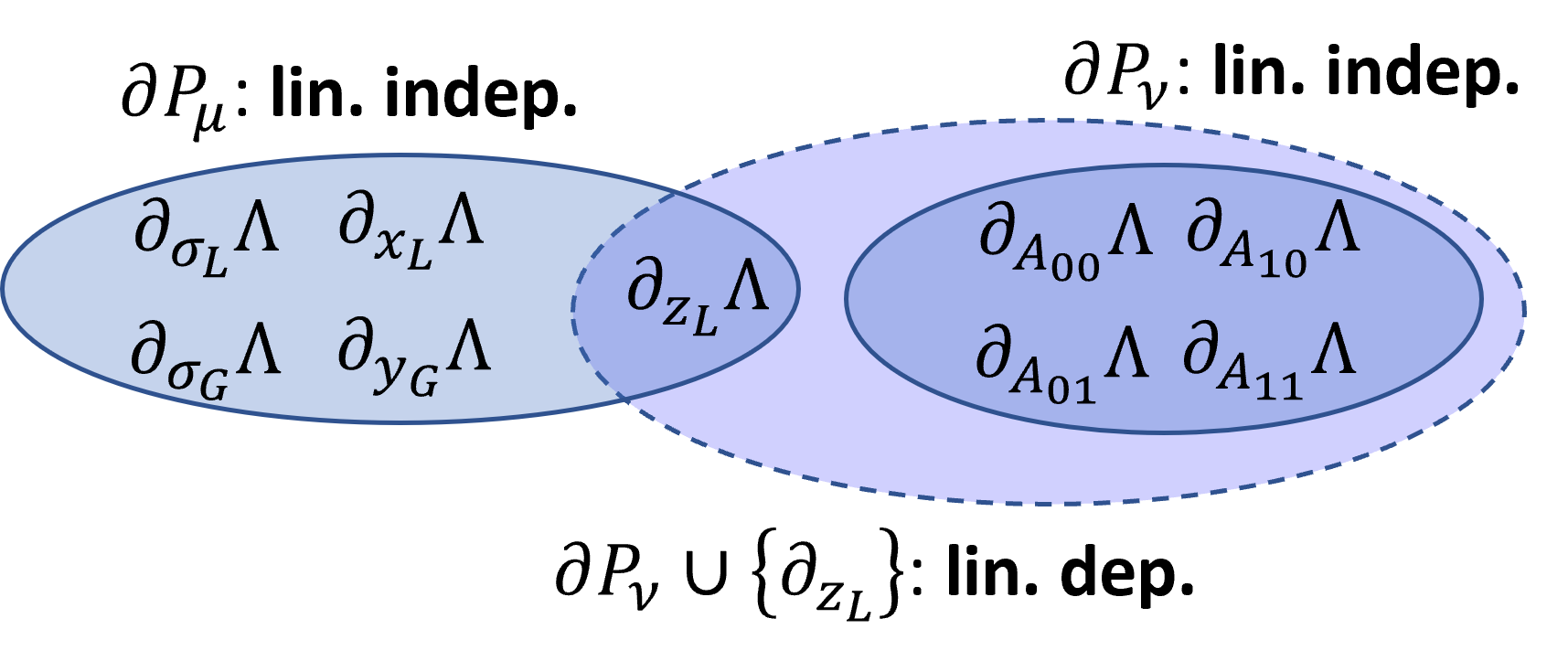}
		\caption[LinearDependencies]{\label{fig:LinearDependencies} Linear dependence (lin. dep.) and linear independencies (lin. indep.) within the set of parameters $\partial P$.}
	\end{figure}

	    The partial derivatives with respect to $\Lambda$ can be calculated straightforwardly:
	    \begin{equation}
		\label{LinearlyIndependentParameters}
		\begin{bmatrix} \partial_{x_L} \\ \partial_{y_G} \\ \partial_{\sigma_G} \\ \partial_{\sigma_L} \end{bmatrix} \Lambda = \begin{bmatrix} \Lambda\left(\frac{r \cos(\theta)-x_L}{\sigma_L^2}\right) \\ \Lambda_{\delta} \left(\frac{r \sin(\theta) - y_G}{\sqrt{(r \sin(\theta) - y_G)^2}}\right) \\ \Lambda_\delta \\ \Lambda\left(\frac{(z-z_L)^2+(r \cos(\theta)-x_L)^2}{\sigma_L^3}\right) \end{bmatrix} ,
	    \end{equation}
	    \begin{equation}
	    \label{LinearlyDependentParameters}
		\begin{bmatrix}  \partial_{z_L} \\ \partial_{A_{mn}} \end{bmatrix} \Lambda = \begin{bmatrix} \Lambda\left(\frac{z-z_L}{\sigma_L^2}\right) \\ \Lambda \left(\frac{\partial z}{\partial A_{mn}}\right) \left(\frac{z-z_L}{\sigma_L^2}\right)  \end{bmatrix} ,
	    \end{equation}
	    where the shorthand $\Lambda_\delta$ represents an analog of $\Lambda$ where the step function is replaced by a delta distribution $H(...)\rightarrow\delta(...)$ (see Eqs.~(16.c) and \eqref{LambdaIntegrandDefinition}). The set 
        $\partial P_{\mu} \setminus \{\partial_{z_L}\}$
         (Eq.~\eqref{LinearlyIndependentParameters}) is linearly independent, but the set 
         $\partial P_{\nu} \cup \{\partial_{z_L}\}$
         (Eq.~\eqref{LinearlyDependentParameters}), which has no dependence on $\theta$, is not necessarily linearly independent (Fig.\ \ref{fig:LinearDependencies}).
	    
	    The linear dependence in $\partial P_{\nu} \cup \{\partial_{z_L}\}$ is seen as:
	    \begin{equation}
	    0=\Lambda \times \left(\sum_{m,n} c_{mn} \frac{\partial z}{\partial A_{mn}} + 1 \right)  \Rightarrow c_{mn}=(n+1) A_{m,n+1} ,
	    \end{equation}
	    where linear independence would require that the only solution is $c_{mn}=0 \Rightarrow A_{m,n}=0$, i.e.\ the trivial solution $T(r,z)=0$. Conversely, every non-trivial form of $T(r,z)\neq 0$ results in linear dependence within the set $P_{\nu} \cup \{\partial_{z_L}\}$. This means that for our TOF calibration procedure to be unique, we must know or fix the value of  $z_L$ independently.
	    
	    Although $z_L$ can not be determined through the TOF spectra, its value could be constrained, which would in turn put a bound on the non-uniqueness problem. We quickly remark how $z_L$ together with the other parameters for $\mu(r,z;P_{\mu})$ describe the starting points of the ion trajectories. If we look back at the Taylor expansion of $T(r,z)$, and explicitly expand it around $z=z_L$:
	    \begin{equation}
	    \begin{aligned}
    		T(r,z)= & A_{00}+A_{01}(z-z_L)+A_{10}r^2\\
            & +A_{11}(z-z_L)r^2+\mathcal{O}(r^4, z^2)
	     \end{aligned} ,
	    \end{equation}
	    we clearly see the invariant shifts leading to the linear-dependence:
	    \begin{equation}
	    \label{SpecificTOF_TaylorExpansion}
	    \begin{cases}
		\begin{matrix} z_L\rightarrow z_L+c \\ A_{00}\rightarrow A_{00}+c \\
		A_{10}\rightarrow A_{10}+c \end{matrix}
		\qquad, c\in\mathbb{R} ,
	    \end{cases}
	    \end{equation}
        As the calibration value for $z_L$ is “dependent” on the potential field, constraints on the electrodes producing the potential field can also be used to constrain $z_L$. If we have an error estimate $\Delta V_f$ on the front electrode $V_f$, the constraint on $z_L$ can be derived through simple error propagation:
	    \begin{equation}
	    \label{eq:VoltageError}
		\Delta z_L \approx \left|\frac{\partial z_L}{\partial V_f}\right| \Delta V_f = \left|\frac{\partial z_L}{\partial T}{\frac{\partial T}{\partial V_f}}\right| \Delta V_f ,
	    \end{equation}
	    so that our calibration procedure is unique up to $\Delta z_L$ or correspondingly $\Delta V_f$.

    \section{Different shape for ionization volume}
		\label{appendix:DifferentIonizationVolume}
	    
	    As a qualitative check for the stability of the reconstruction, we will assume a different form for the ionization volume in Eq.~(16a), which is more typical of crossed-beam geometry: we keep $I_L(x,y,z)$ as is, but change $I_G(x,y,z)$ from the shape of a planar jet to a a collimated cylindrical jet:
	    \begin{equation}
	    \label{eq:CylindricalGasProfile}
		\begin{aligned}
		I_G(x,y,z) &= \frac{1}{2 \sigma_G} H(\sigma_G - \sqrt{(x-x_L)^2 + (y-y_G)^2}) . \\
		\end{aligned}
	    \end{equation}
	    We then re-perform the entire reconstruction procedure including calibration to yield KER spectra analogous to Fig.\ 6. The found fit parameters are shown in Table \ref{table:CylinderFitParameters}.
	\begin{table}[b!]
		\centering
		\caption[table:CylinderFitParameters]{\label{table:CylinderFitParameters} Fit parameters of the ionization volume according to Eqs.~\eqref{eq:CylindricalGasProfile} and Eq.~(16a), found through the calibration procedure.}
    \begin{ruledtabular}
		\begin{tabular}{ l | c | c } 
			Description (\& units)  & Fit value & Error (1$\sigma$) \\
			Variable in Eqs.~(14), (16a) &  &   \\ 
			
			\hline\hline
			x-centre of laser ($\upmu$m)   & 8374.76  & 0.12( + 350)\footnotemark[1] \\
			\quad 0 $\leq x_L \leq$ 20000  &          &                    \\		
			z-centre of laser (mm)         & 0        & fixed\footnotemark[2]\\
			\quad 0 $\leq z_L < ~$5        &          &                    \\	
			width of laser ($\upmu$m)      & 44.37    & 0.11               \\	
			\quad 0 $< \sigma_L$           &          &                    \\			
			\hline
			y-centre of gas jet (mm)       & 0.121    & 0.057              \\
			\quad 0 $\leq y_G < ~$5        &          &                    \\		
			width of gas jet (mm)          & 1.862    & 0.057              \\
			\quad $\sigma_G < ~$2.5        &          &                    \\			
			\hline	
			photopeak undersampling        & 1.862    & 0.029              \\
			\quad 0 $< c_\delta$           &          &                    \\					
			baseline decay (ns)            &  9.53    & 0.24               \\
			\quad 0 $< \tau_g$             &          &                    \\				
			baseline factor                & -0.0222  & 0.0009             \\
			\quad $c_\tau \leq$ 0          &          &                    \\				
		
		\end{tabular}
    \end{ruledtabular}
    \footnotetext[1]{Additional error from the uncertainty of the electrode voltages in Eq.~\eqref{eq:VoltageError} (see Appendix \ref{appendix:CalibrationUncertainties}).}
    \footnotetext[2]{The assumption $z_L=0$ was done for practical reasons, but otherwise did not significantly affect the fitting.}
		
	\end{table}
	    
       This alternative profile results in KER spectra which are near-identical to the one in Fig.\ 6, and the corresponding fit parameters remain well within the errors of the fitted parameters in Table I. This shows that the calibration and reconstruction are stable with regard to small changes in the shape of the ionization volume after optimization (compare Eq.~(16c) with Eq.~\eqref{eq:CylindricalGasProfile}).

\bibliography{supplementaryReferences}